\documentclass[journal,onecolumn,12pt]{IEEEtran}
\usepackage{amsthm,graphicx}
\usepackage{amsmath}
\usepackage{amssymb}
\usepackage{cite}

\theoremstyle{plain}
\newtheorem{thm}{\protect\theoremname}
\theoremstyle{definition}
\newtheorem{defn}{\protect\definitionname}
\theoremstyle{remark}
\newtheorem{rem}{\protect\remarkname}
\theoremstyle{plain}
\newtheorem{cor}{\protect\corollaryname}
\theoremstyle{plain}
\newtheorem{lem}{\protect\lemmaname}

\providecommand{\corollaryname}{Corollary}
\providecommand{\definitionname}{Definition}
\providecommand{\lemmaname}{Lemma}
\providecommand{\remarkname}{Remark}
\providecommand{\theoremname}{Theorem}
 
\makeatletter

\providecommand{\tabularnewline}{\\}

\begin{document}

\title{On the Energy Complexity of LDPC Decoder Circuits\thanks{\noindent Submitted
for publication on February 25th, 2015.  Presented in part
at the 2014 IEEE North American School of Information Theory,
June 18--21, Toronto, Canada.} \author{Christopher Blake and Frank R. Kschischang\\Department of Electrical \& Computer Engineering\\University of Toronto\\ \texttt{\small christopher.blake@mail.utoronto.ca} \texttt{\small frank@comm.utoronto.ca}}}
\maketitle
\begin{abstract}
It is shown that in a sequence of randomly generated bipartite configurations with number of left nodes approaching infinity, the probability
that a particular configuration in the sequence has a minimum bisection width proportional to the number of
vertices in the configuration approaches $1$ so long as a sufficient condition on the node degree
distribution is satisfied. This graph theory result implies an \emph{almost sure} $\Omega\left(n^{2}\right)$ scaling rule for the energy of capacity-approaching LDPC decoder circuits that directly instantiate their Tanner Graphs and are generated according
to a uniform configuration model, where $n$ is the block length of the code. For a sequence of circuits that have a full set of check nodes but do not necessarily directly instantiate a Tanner graph, this implies an $\Omega\left(n^{1.5}\right)$ scaling rule. In another theorem, it is shown that
\emph{all} (as opposed to almost all) capacity-approaching LDPC decoding circuits that directly implement their Tanner graphs must have energy that scales as $\Omega\left(n\left(\log n\right)^{2}\right)$. These results further imply scaling rules for the energy of LDPC decoder circuits as a function of gap to capacity.

\end{abstract}

\section{Introduction}

Low density parity check codes are a class of codes first introduced
by Gallager in \cite{GallagerLDPC}. This paper finds fundamental lower
bounds on the energy of VLSI implementations of capacity-approaching LDPC decoders. Central to the construction and analysis of LDPC codes
is the randomly generated Tanner graph with a given degree distribution.
A widely used method of analysis involves analyzing an ensemble of LDPC codes whose
Tanner graphs are generated according to some distribution.
It has been shown that there exist degree distributions that result
in LDPC codes and decoders that can get arbitrarily close to capacity
for an erasure channel \cite{OswaldCapacityApproachingLDPCErasureChannel}. The first main result of this paper is an "almost-sure" scaling rule for the energy of capacity-approaching LDPC decoders whose Tanner graphs are generated according to a uniform configuration model. The second main result of this paper is a scaling rule for the energy of all, as opposed to almost all, capacity-approaching LDPC decoders. What we mean by an "almost sure" and "sure" scaling rule will be made more precise later in the paper. 

To find energy-complexity lower bounds on a class of algorithms a
computation model is needed. We use a standard circuit model that was
first presented by Thompson in \cite{Thompson}. In this model, 
we consider the energy of a circuit implementation
of an algorithm to be the area of the circuit multiplied by the number
of clock cycles required to execute the algorithm. We will give a
more detailed discussion of this model later in the paper. The authors
of \cite{groverFundamental} used the Thompson model to analyze the energy complexity of all
decoding algorithms by showing that
as the target block error probability approaches $0$, the total energy must approach
infinity. In \cite{BlakeKschischangFundamentalLowerBoundArxiv} the authors showed that any fully-parallel
decoding scheme that asymptotically has block error probability
less than $\frac{1}{2}$ must have energy complexity which scales
as $\Omega\left(n\sqrt{\log n}\right)$. These results, though general,
do not suggest the existence of any decoder implementations that reach
these lower bounds. In this paper, we in particular show that the energy
of LDPC decoding schemes that directly-implement their Tanner graphs cannot reach
the $\Omega\left(n\sqrt{\log n}\right)$ energy lower bound, and in fact must
have energy that scales at least as $\Omega\left(n\left( \log n\right)^2\right)$.

We begin the paper in Section \ref{sec:Background} with a discussion
of the graph theory used in the paper, and we also discuss some prior
work that reaches similar conclusions to our paper. Then, in Section
\ref{sec:Definitions-and-Main-Lemmas} we introduce graph theory definitions
and the circuit model that we will use. We also present some important
lemmas that will be used in our theorems. Then, in Section \ref{sec:Main-Theorem},
after defining some properties of node degree distributions, we present
the main theorem which shows that almost all LDPC Tanner graphs have
minimum bisection width proportional to the number of vertices. We
proceed to show how this theorem allows us to find scaling laws for
the energy of directly-implemented LDPC decoders in Section \ref{sec:Almost-Sure-Bounds-on-LDPC-Circuits}.
The results presented in these sections are true for almost all LDPC
decoders (\emph{i.e.,} for a set of decoders with probability approaching
one), but it is not clear whether there is a set of LDPC decoders
of probability approaching $0$ that can approach capacity. Thus,
in Section \ref{sec:Bounds-for-All} we present a theorem that relates
the number of edges and vertices in a graph to the area of its circuit
instantiation to show a scaling rule that is applicable to any LDPC
decoding algorithm that approaches capacity. This results in a \emph{sure}
as opposed to \emph{almost sure} scaling law for the energy per iteration
of a directly-instantiated LDPC decoder of $O\left(n\left(\log n\right)^{2}\right)$.

\section{Background\label{sec:Background}}

\subsection{Related Work on LDPC Scaling Rules}

There are some results on fundamental limits on wiring complexity
of LDPC decoders. In particular, in
\cite{GanesonGroverLDPCLowerBound}, the authors assume that the average wire
length in a VLSI instantiation of a Tanner graph is proportional to
longest wire in an asymptotic sense, and that the longest wire is
proportional to the diagonal of the circuit upon which the LDPC decoder
is laid out. The implication of these assumptions is an $\Omega\left(n^{2}\right)$
scaling rule for the area of directly-implemented LDPC circuits, which
is the same result of this paper. However, these assumption are taken
as axioms without being fully justified; there certainly can exist
bipartite Tanner graphs that can be instantiated in a circuit without
such area. The result of this paper suggests that, in fact, the $\Omega\left(n^{2}\right)$
scaling rule is justified for \emph{almost all } VLSI instantiations
of LDPC Tanner graphs as the block length of these LDPC codes grow
large, where the Tanner graphs are generated from a uniform configuration
model and a sufficient condition on the node degree distributions
is satisfied. This scaling rule is an implication of the main theoretical
contribution of this paper: a result in random graph theory that we present as Theorem \ref{thm:MainTheorem}.
In addition to this, we provide a super-linear energy
scaling rule for \emph{all} directly-implemented LDPC decoders, even
if the Tanner graph of such decoders is not generated according to
the uniform configuration model.

\subsection{Related work on Graph Theory}

In graph theory, there are a number of results that study the minimum-bisection
width of graphs. Often this work looks at a graph's Laplacian, which is a matrix equal to the difference in the graph's degree matrix and adjacency matrix. In \cite{FiedlerBisectionWidthLowerBound} a graph's Laplacian is analyzed and it is shown that the second largest eigenvalue, $\lambda_2$, can be used to find a lower bound of $\frac{\lambda_2 n}{4}$ on the graph's minimum bisection width. In \cite{BezrukovSpectralLowerBounds}, the
authors find some bounds on the bisection width of graphs that are related to this $\lambda_2$ value.  The authors in \cite{DiazBisectionWidthBoundsRegularGraphs}
provide almost sure upper bounds for the bisection width of randomly
generated regular graphs. Our result does not consider the second greatest eigenvalue of the Laplacian of a graph to bound the minimum bisection width. Instead, we use a unique purely combinatorial approach to reach our almost sure lower bounds. Furthermore, our analysis is of random bipartite graphs, as opposed to random regular graphs. As well, our result makes only weak assumptions on the node degree distribution to get our lower bound, without requiring a degree-regularity assumption. The generality of the result allows us to apply the theorem to find a scaling rule for the area of almost all capacity-approaching directly-implemented LDPC decoding circuits.

\section{\label{sec:Definitions-and-Main-Lemmas}Definitions and Main Lemmas}

\subsection{Graph Theory Definitions}

The main result of our paper involves the minimum bisection
width of a graph. The minimum bisection width is a property of any
graph. A bisection is a set of edges that once removed divides the
graph into two subgraphs that have the same number of vertices. A
formal definition is given below.

\begin{figure} 
\centering
\raisebox{3ex}{\includegraphics{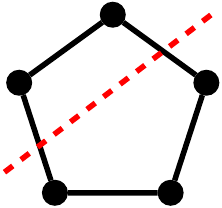}}~\includegraphics{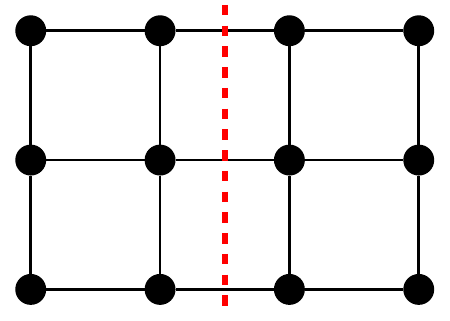}
\caption{Example of two graphs with a minimum bisection labelled. Nodes are
represented by circles and edges by lines joining the circles. A dotted
line crosses the edges of each graph that form a minimum bisection.}
\label{minimumBisectionWidth}
\end{figure}
\begin{defn}
Consider a graph $G$ with vertices $V$ and edges $E$. Let $E_{s}\subseteq E$
be a subset of the edges. Then $E_{s}$ \emph{bisects} $G$ if removal of
$E_{s}$ cuts $V$ into unconnected sets $V_{1}$ and $V_{2}$ in which
$\left|\left|V_{1}\right|-\left|V_{2}\right|\right|\le1$. A \emph{minimal
bisection} is a bisection of a graph whose size is minimal over all
bisections. The \emph{minimum bisection width} is the size of a minimal bisection. 
\end{defn}
Generally speaking, finding the minimum bisection width of a graph
is a difficult problem (it is in fact NP-Complete \cite{Garey1976237}). The diagram
in Fig. ~\ref{minimumBisectionWidth} shows minimal bisections of
a few simple graphs. Associated with a bisection $E_{s}$ of a graph
$G$ are two unconnected graphs $G_{1}=\left(V_{1},E_{1}\right)$
and $G_{2}=\left(V_{2},E_{2}\right)$ induced by the bisection. We
will refer to the set of vertices $V_{1}$ and $V_{2}$ each as a
\emph{bisected set of vertices} \emph{induced by a bisection} or,
more compactly, a \emph{bisected set of vertices}, where the association
with the particular bisection is to be implicit.

Note that in this paper we will often consider dividing the vertices
of a subset into two disjoint sets $V_{1}$ and $V_{2}$ in which
$\left|\left|V_{1}\right|-\left|V_{2}\right|\right|\le1$. For convenience
of discussion, we call this process \emph{dividing the vertices in
half. } We make particular note of this to avoid in every case having
to distinguish between if the cardinality of the set of vertices in
question is even or odd.

\subsection{Circuit Model}

\begin{figure} \centering \includegraphics{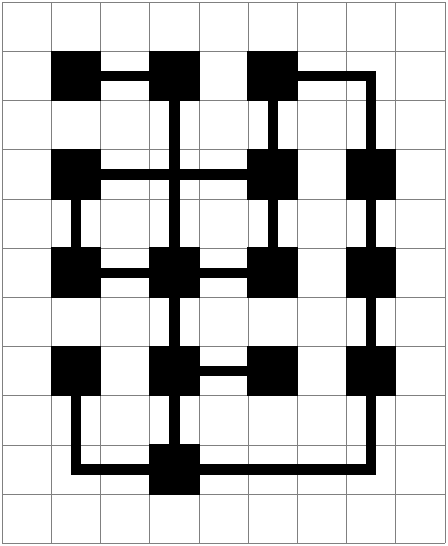} \caption{Diagram of a possible VLSI circuit. Grid squares that are fully filled in represent computation nodes and the lines between them represent wires. } \label{fig:circuit} \end{figure}

Central to our discussion is the
relation between minimum bisection width of a graph and the area (and thus energy)
of a circuit that implements that graph. Our discussion applies directly
to LDPC decoders, and within our model we must define an LDPC decoder,
as well as a more general circuit. In this paper, the definition of
a \emph{circuit} is adapted from Thompson \cite{Thompson} and is
considered to be a mathematical object consistent with the following circuit axioms. This model was also used in \cite{groverFundamental}
to find bounds on the energy complexity of encoding and decoding algorithms.
We also provide a diagram of an example circuit in Figure \ref{fig:circuit}.

\begin{itemize}
\item A circuit is a collection of nodes and wires laid out on a planar
grid of squares. Each grid square can be empty, can contain a \emph{computational
node} (sometimes referred to more simply as a node), a \emph{wire},
or a \emph{wire crossing}. A circuit also has some special nodes called
\emph{input nodes} and also \emph{output nodes}. The purpose of a circuit is to compute a function $f:\left(0,1\right)^{n}\rightarrow\left(0,1\right)^{k}$. Such a circuit is said to have $n$ inputs and $k$ outputs. The computation
is divided into $\tau$ clock cycles.  The inputs into a computation are to be loaded into
the input nodes, and the outputs are to appear in the output nodes during some set clock cycle of the computation.
\item Each grid square has width $\lambda_{\mathrm{w}}$, known as the \emph{wire
width }and thus has area $\lambda_{\mathrm{w}}^{2}$. It is in this parameter
that this circuit model subsumes different VLSI implementation techniques.
In real circuits, this parameter may be a value like $14$ nanometers.
Our concern in this paper is not what this value is, but rather in
providing scaling rules in terms of the VLSI implementation technology
used.
\item The \emph{computational nodes} are the ``computing'' parts of the
circuit. A node has at most $4$ bidirectional wires connected to it, which are used to feed in bits into the node and feed out the bits computed by the node. Each node is 
capable of computing a fixed function of the bits fed into it by the wires connected to them during each clock
cycle. In particular, a node with $f\le4$ wires leading into it can
compute any function $g:\left\{ 0,1\right\} ^{f}\rightarrow\left\{ 0,1\right\} ^{f}$. However,
a computational node is restricted to only be able to compute the
same function at each clock cycle. We note, of course, that the output
of a particular node could change with each clock cycle because, in
general, the inputs into the function could change with each clock
cycle.
\item The \emph{wires} are the ``communication'' part of a circuit. Wires
in a circuit are connections between computational nodes, and are
assumed in our model to be bidirectional. At each clock cycle a wire
can carry one bit in each direction. The bits communicated are an
output of the function computed by the computational node to which
the wire is connected. A wire can be placed in a grid square in a
way that connects one edge of the grid square to some other edge.
Thus, grid squares containing wires can be connected to form a wire
leading from one node to another node.
\item An \emph{input node} is a special node in the circuit. In addition
to being able to compute any fixed function mapping its $f\le4$ inputs
to its $f\le4$ outputs, this node is also given an input bit into the circuit. In general, at each clock cycle an input node can have as its input a new input into the function. Thus, we say that inputs, in general, can be \emph{serialized}; that is, they can be injected into the circuit at different clock cycles of the computation. Usually it is assumed that
the inputs into an input node are chosen from the set $\left\{ 0,1\right\} $;
however, sometimes (especially for the purpose of lower bound) we
can assume that the inputs into an input node are chosen from a larger
set of values. In \cite{BlakeKschischangFundamentalLowerBoundArxiv} it was assumed that an input node
that is attached to $f$ wires can compute any function $g:\left\{ 0,1,?\right\} \times\left\{ 0,1\right\} ^{4}\rightarrow\left\{ 0,1\right\} ^{4}$,
\{i.e., the node can perform any function of its $4$-bit input from
the wires connecting to it, as well as its input, taken from the symbols
$\left\{ 0,1,?\right\} $, where in the case of this assumption $?$
is considered an erasure symbol. In our analysis we can assume that an erasure is a valid input as well, however this is not a central assumption of this paper and the results apply to inputs being taken from the set $\left\{ 0,1\right\} $.
\item An \emph{output node} is another special node in a circuit. It is
permitted to, like any other node, compute any function of its inputs,
but it is given an additional output. Thus, in the case
of an output node with $f\le4$ wires leading from it, the output node
can perform any function $g:\left\{ 0,1\right\} ^{f}\rightarrow\left\{ 0,1\right\} ^{f+1}$
where one of the bits in the output is distinguished as an \emph{output
bit}. The output node is required
to hold in its output bit some circuit output during set clock cycles. In a fully parallel computation the output node is required to hold one output bit of the computation at the end of the computation, but in general the outputs may be serialized, and one output node can be responsible for outputting a number of the outputs of the computation, where each output has a specified clock cycle during which it is to appear.
\item A \emph{wire crossing} in a circuit is a grid square that contains
two wires that ``cross'' each other. An example of a circuit with
computational nodes, wires, and a wire crossing is given in Fig. \ref{fig:circuit}.
\item The \emph{normalized area} of a circuit is the number of grid squares
occupied, and it is denoted with the symbol $\overline{A}$. The number
of grid squares occupied with nodes/wires is the \emph{normalized
area of the nodes/wires} of the circuit, and is denoted $\bar{A}_{\mathrm{n}}$/$\bar{A}_{\mathrm{w}}$.
Thus, the actual area of the circuit is $A=\lambda_{\mathrm{w}}^{2}\bar{A}$
and the area of the nodes/wires are defined similarly by multiplying
the normalized value by $\lambda_{\mathrm{w}}^{2}$, the area of a unit grid
square.
\item The \emph{energy} of a computation is proportional to the product
of the area of the circuit, times the number of clock cycles. Real VLSI circuits are made of conducting material laid out essentially flat; thus, in our model, we say that the
capacitance of a circuit is proportional to its area. A circuit works by, at every
clock cycle, charging or discharging its wires. It is thus assumed
that the energy of a computation is proportional to $\frac{1}{2}CV_{dd}^{2}\tau$
where $C=C_{\text{unit-area}}A$. Thus, we can denote the energy of
a computation as $E_{\text{comp}}=\xi_{\text{tech}}A_{\text{c}}\tau$
where $\xi_{\text{tech}}=\frac{1}{2}C_{\text{unit-area}}V_{dd}^{2}$
is a constant that varies depending on the technology used to implement
the circuit. For decoder circuits we often denote the energy of computation
as $E_{\mathrm{dec}}$ where the subscript indicates the type of computation
performed by the circuit under consideration.
\end{itemize}
Note that the restriction that each node has at most four inputs and
four outputs is somewhat arbitrary; it is also arbitrary that each
node is permitted to compute any function of its inputs all at the
same cost. In real VLSI implementations it may be that an arrangement
of transistors can compute some functions more efficiently than others.
However, our model does not consider what gains could be made if certain
functions are cheaper in an energy sense to compute. On the other hand, the model subsumes the interconnection complexity of the inputs
of the function to their outputs. In the field of error control codes,
this interconnection complexity has been shown to be a significant
factor in the energy of a computation in, for example, \cite{1023755,groverOverDesigning}.

\subsection{Relationship Between Circuit Model and Graphs}

This paper analytically characterizes a relationship between the energy
of LDPC decoders as a function of block length and gap to capacity. To understand this we must first define what is meant
by an LDPC decoder implemented according to the Thompson VLSI model.
To understand this we must first understand the connection between
a circuit and the graph corresponding to a circuit.

 Note that a circuit is a collection of nodes connected by wires. Each of the
computational nodes of a circuit can be thought of as the vertices
of a graph, $G=\left(V,E\right)$. The wires of a circuit correspond
to the edges of a graph. In particular, two vertices $v_{1}$ and $v_{2}$
are connected in the graph $G$ by an edge if and only if there is
a wire connecting the two computational nodes that correspond
to $v_{1}$ and $v_{2}$. Thus, any circuit can be considered a graph.
As well, any graph can be implemented as a circuit (although of course
there may be many ways to implement a particular graph on a circuit).
Note that although a circuit, according to our model, must be planar,
since we also allow wire crossings, any graph can be implemented,
though it may be that more complex graphs require far more circuit
area.

Note that saying that a circuit has a corresponding graph is a slight
abuse of terminology: a graph, according to common definitions, does
not allow for two edges between the same nodes, but obviously two
computational nodes are permitted to have two or more wires connecting
them. More precisely, we mean that a circuit has a corresponding multi-graph.
However, for the sake of simplicity we simply call a circuit's corresponding
multigraph a graph, and we hope that this does not cause confusion
for the reader.

Sometimes in our discussion we may want to refer not to a particular
node of the circuit (corresponding to the node of a graph), but rather to
the nodes associated with a subcircuit, which leads to the following
definition.
\begin{itemize}
\item A \emph{subcircuit} is a circuit corresponding to a subset of nodes
of the graph and the wires connecting them. In particular, it is the circuit induced by deleting all wires not connecting the nodes of interest and by deleting all the other nodes in the graph. Any subcircuit has associated with it both
\emph{internal wires} (the wires connecting the nodes of this circuit)
and also \emph{external wires}, the wires leading from nodes within
the subcircuit to nodes from outside the subcircuit. Note that the
notion of a subcircuit corresponds to a particular subgraph of the
graph of the circuit. In the language of graph theory \cite{douglasb.west2001}, we can say that a subcircuit with
computational nodes corresponding to some subset of $V'\subseteq V$
corresponds to the subgraph \emph{induced} by the vertices in $V'$.
Note that any subset of the computational nodes of a graph induces
a subcircuit and also a subgraph of the circuit's graph.
\end{itemize}

\subsection{LDPC Decoders}

An LDPC code is a linear code first invented by Gallager in \cite{GallagerLDPC}.
All linear codes can be specified by a parity check matrix. Central to
the construction LDPC codes is the  \emph{Tanner graph} of the code corresponding to a parity check matrix of the code. A Tanner
graph is a bipartite graph. Thus, such a graph has two \emph{partite
sets}, or sets of unconnected vertices which are referred to as the
\emph{check nodes} and the \emph{variable nodes}. An $\left(n,k\right)$ LDPC
code has associated with it a Tanner graph with $n$ variable nodes
and at least $n-k$ check nodes (we say at least because it may be
that some of the linear constraints induced by the check nodes are
not linearly independent). The $n$ variable nodes correspond to the
$n$ symbols of a block length $n$ codeword in the LDPC code. A codeword
$c\in\left\{ 0,1\right\} ^{n}$ is in the LDPC code generated by a
Tanner graph if, for each check node in the Tanner graph of the code,
the $\mod 2$ sum of the values of the variable nodes to which they
are connected is $0$. The association of a set of linear constraints
with a Tanner graph leads to natural and very efficient methods of
decoding that exploit the sparse nature of the Tanner graph. 

An LDPC decoding algorithm associated with a Tanner graph is a message-passing procedure. Each variable node is thought conceptually to be
connected to their check nodes, and each check node correspondingly
to their variable nodes. In general, a variable node has as its inputs
a message passed to it from each of the check nodes to which it is
connected, as well as the output of a noisy channel. A variable node,
in general, is able to compute any function of these inputs and pass
the outputs of this computation to its adjacent check nodes. The check
nodes are similarly allowed to compute any function of their inputs
(which will be in general the outputs of the variable nodes to which
they are connected). An \emph{iteration }of an LDPC decoder is one
instance of this procedure: the variable nodes computing a function
that is then passed to the check nodes, and then the check nodes computing
a function of these messages and passing the output of these functions
back to the variable nodes to which they are connected. A good LDPC
decoding algorithm should choose these functions well, so that, at
the end of a certain number of clock cycles $\tau$, the variable
nodes hold within them an estimate of the original input into a noisy
channel. In the most general case, we allow the check and variable
nodes to compute different functions of their inputs during different
iterations (i.e., the function they compute in general may vary in
time). Gallager discussed a variety of these message passing procedures
in \cite{GallagerLDPC}.

To instantiate an LDPC decoding algorithm in a circuit, we consider
two possible paradigms, a \emph{directly-implemented} technique in
which the Tanner graph of an LDPC code is directly instantiated in
some sense by the circuit, and a \emph{complete-check node serialized} technique, in which
the Tanner graph is not necessarily directly implemented, but there are subcircuits in the graph corresponding to each check node and an LDPC message passing procedure is performed.

A directly-instantiated LDPC decoder can be thought of as a circuit
that has a graph that is
an implementation of a Tanner graph of the underlying LDPC code. To
be precise, we will use terminology borrowed from graph theory regarding
the subdivision of a graph.

\begin{defn} Suppose a graph has an edge, $e$, connecting vertices
$v_{1}$ and $v_{2}$. Then a \emph{subdivision
}of edge $e$ in a graph is a process that takes the graph $G$ and
forms a new graph $G'$ with an additional vertex $v'$ and two additional
edges connecting $v_{1}$ and $v_{2}$ to $v'$ by replacing $e$
with two edges. A \emph{subdivision} of a graph $G$ is a graph obtained
by the successive subdivisions of edges in the graph.
\end{defn}

If a graph $G$ has a subgraph that is a subdivision of a graph $G'$,
then we say that the graph $G$ \emph{contains} graph $G'$.
This leads to an important lemma that will allow us to connect bounds
on graph properties of a Tanner graph to the area of directly-implemented
LDPC decoders.

\begin{defn} \label{Definition:DirectlyImplementedLDPCDecoder}
A \emph{directly-implemented LDPC decoder} is a circuit associated
with an LDPC code with a Tanner graph $\mathcal{T}$.
Consider the graph associated with the circuit. Then a circuit is
a directly-implemented LDPC decoder if its graph contains $\mathcal{T}$. 
\end{defn}
This means that a circuit is a directly-implemented LDPC decoder if
there are subcircuits corresponding to each variable node and edges
leading from these ``black boxes'' that connect to subcircuits that
correspond to the check nodes of the Tanner graph.

Associated with any graph $G$ is a quantity that we will call the
\emph{minimum area of a circuit implementation of $G$, }or, to be
more concise, \emph{the area of $G$. }The area of a graph $G$ is
the circuit with corresponding graph $G$ with the minimum number
of grid squares occupied. We denote this quantity as $A_{\min}\left(G\right)$.
\begin{lem}
\label{lem:AMinLemma} If a graph $G$ contains a graph $G'$, then
$A_{\min}\left(G\right)\ge A_{\min}\left(G'\right)$.\end{lem}
\begin{rem}
This is a very intuitive idea. If a graph contains another graph,
then naturally one would regard the original graph as ``larger''
in some sense then the graph that it contains. This notion will be
used to connect a bound on the area of a circuit implementing the
Tanner graph of an LDPC code to a bound on directly-implemented
LDPC decoders.\end{rem}
\begin{IEEEproof}
Suppose that $A_{\min}\left(G\right)<A_{\min}\left(G'\right)$. Consider
the circuit with minimal area that implements $G$. We can use this
circuit to construct a circuit for $G'$ with area less than $A_{\min}\left(G'\right)$,
resulting in a contradiction. Since $G$ contains $G'$, there is
a subgraph of $G$ that is a subdivision of $G'$. Consider the subcircuit
associated with that subgraph. Clearly, this subcircuit has area less
than or equal to $A_{\min}\left(G\right)$. Delete those nodes of this
subgraph that correspond to subdivisions of edges of $G'$. On a circuit,
this corresponds to replacing a computational node with merely a wire.
This process does not change the area of this subgraph, and it will
result in a circuit for $G'$ less than $A_{\min}\left(G'\right)$,
a contradiction.
\end{IEEEproof}
There is a key result attributed to Thompson \cite{Thompson} that relates a graph's minimum bisection width to the area of a circuit implementing that graph,
presented in the following lemma.
\begin{lem}
\label{lem:ThompsonLemma}If a graph has minimum bisection width $\omega$,
then the area of a circuit implementing this graph is lower bounded
by
\[
A_{\mathrm{c}}\ge\frac{\lambda_{\mathrm{w}}^{2}\omega^{2}}{4}.
\]
\end{lem}
\begin{IEEEproof}
See Thompson \cite{Thompson} for a detailed proof. 
\end{IEEEproof}
Currently, our definition of a directly-implemented LDPC decoder
subsumes many practical implementations of LDPC decoding algorithms,
but in practice circuits can be implemented that perform an LDPC decoding algorithm and do not directly instantiate
the Tanner graph of the code. This thus motivates the following definition of a more general type of LDPC decoder.
\begin{defn}
An $\left(n,k\right) $ \emph{complete-check-node LDPC decoder} associated with Tanner graph $\mathcal{T}$ is a circuit
 with $n$ separate subcircuits each corresponding to a variable node in $\mathcal{T}$ and one subcircuit corresponding to each check node in $\mathcal{T}$. During one iteration a message must be passed from each variable-node subcircuit to each
adjacent check-node subcircuit, and also from each check-node subcircuit to each adjacent variable-node subcircuit. To be precise, the check-node subcircuits that are adjacent to a variable-node subcircuit are those check-node subcircuits that correspond to check nodes in $\mathcal{T}$ that are adjacent to the variable node that corresponds to the variable-node subcircuit of interest.  The variable-node subcircuits that are adjacent to a check-node subcircuit are defined similarly.
\end{defn}
Note that for such a circuit we do not require that a wire exists in the circuit for each edge in the Tanner graph. Thus, it is possible that a complete-check-node LDPC decoder can use the same wire multiple times, but in different clock cycles to communicate information during an iteration.

Our results rely on the evaluation of some limits, which we present
as lemmas below.
\begin{lem}
\label{lem:OverallLimitLemma}Suppose $P\left(n\right)=O\left(n^{k}\right)$
for some $k>0$ and is positive for sufficiently large $n$, and there is a sequence $n_1, n_2, \ldots$ that increases without bound. Then:
\end{lem}
\begin{eqnarray*}
\lim_{i\rightarrow\infty}P\left(n_i\right)\exp\left(-n_i f\left(n\right)\right) & = & 0\text{ if}\\
\lim_{n\rightarrow\infty}f\left(n\right) & > & 0.
\end{eqnarray*}

\begin{IEEEproof}
Since $\lim_{n\rightarrow\infty}f\left(n\right)>0$ and the sequence $n_i$ increases without bound, then for sufficiently
large $i$, $f\left(n_i\right)>c$ for some $c>0$ (in particular for
any $c$ strictly less than the value of the limit). Then, for sufficiently
large $i$, 
\[
P\left(n_i\right)\exp\left(-n_i f\left(n_i\right)\right)\le P\left(n_i\right)\exp\left(-cn_i\right).
\]
Clearly, $\lim_{i\rightarrow\infty}P\left(n_i\right)\exp\left(-cn_i\right)=0$
and because $P\left(n\right)$ is positive for large enough $n$,
$P\left(n_i\right)\exp\left(-n_if\left(n_i\right)\right)>0$ for large enough
$i$. The limit thus follows from the squeeze theorem. \end{IEEEproof}
\begin{lem}
\label{lem:MFactorialNFactorialUpperBoundLemma}For any two positive
integers $m$ and $n$ in which 
\begin{equation}
m+n\le Y\label{eq:mPlusnBound}
\end{equation}
 for an integer $Y>0$ where $Y\le Z$ and both $m\le Z$ and $n\le Z$,
\begin{equation}
m!n!\le Z!\left(Y-Z\right)!.\label{eq:ZfactorialYMinusZfactorialBound}
\end{equation}
\end{lem}
\begin{IEEEproof}
Since $m+n\le Y$, then $n=Y-m$ surely maximizes the product $m!n!$
(regardless of any additional restriction on $n$). Suppose a possible
choice of $m=Z-c$ and $m=Y-Z+c$, for some $c>0$ in which $Y-Z+c\le Z$. We divide $Z!\left(Y-Z\right)!$ by the quantity $\left(Z-c\right)!\left(Y-Z+c\right)!$ and show that this quantity is greater than or equal to $1$, meaning that $Z!\left(Y-Z\right)!$ maximizes the product:
\begin{align*}
&\frac{Z!\left(Y-Z\right)!}{\left(Z-c\right)!\left(Y-Z+c\right)!}=\\
&\frac{Z\left(Z-1\right)\ldots\left(Z-c+1\right)}{\left(Y-Z+c\right)\left(Y-Z+c-1\right)\ldots\left(Y-Z+1\right)}
\end{align*}

Note that the numerator and denominator have precisely $c$ terms.
Since $Z\ge Y-Z+c$ the terms in the numerator are strictly greater
than a corresponding term in the denominator, unless $Y-Z+c=Z$, but
of course in this case the product is merely equal to the upper bound
in (\ref{eq:ZfactorialYMinusZfactorialBound}). 
\end{IEEEproof}

\section{Main Theorem\label{sec:Main-Theorem}}

Our main theorem is fundamentally graph-theoretic in nature and applies
to graphs generated according to a standard uniform random configuration
model. We present this theorem in a general form and then specialize
it to create an ``almost sure'' scaling rule for capacity-approaching
LDPC codes.

Consider the set of bipartite graphs $G=\left(V_{L}\amalg V_{R},E\right)$
in which $\left|V_{L}\right|=n$, $\left|V_{R}\right|=m$, and with
left node degree sequence $\Lambda=\left(\lambda_{1},\lambda_{2},\ldots,\lambda_{n}\right)\in\left(\mathbb{N}\right)^{n}$
and right node degree sequence $P=\left(\rho_{1},\rho_{2},\ldots,\rho_{m}\right)\in\left(\mathbb{N}\right)^{m}$.
In other words, for a particular graph in this set, $\lambda_{i}$
is the degree of $v_{i}\in V_{L}$, the $i$th left node in the graph,
and $\rho_{i}$ is the degree of $r_{i}\in V_{R}$, the $i$th right
node in the graph. Without loss of generality, assume that the degree
sequences are ordered, \emph{i.e.} that $\lambda_{1}\le\lambda_{2}\le\ldots\le\lambda_{n}$
and $\rho_{1}\le\rho_{2}\le\ldots\rho_{m}$, and also, without loss
of generality, assume $n\ge m$. Denote this set $\mathcal{G}\left(\Lambda,P\right)$.
Note that the number of edges in each particular graph in $\mathcal{G}\left(\Lambda,P\right)$
is $\left|E\right|=\sum_{i=1}^{n}\lambda_{i}=\sum_{i=1}^{m}\rho_{i}$.

For convenience of counting, we will consider not the set of graphs with a particular degree sequence, but rather the set of \emph{configurations} with this degree sequence. We can associate
each node in a graph with a number of sockets equal to its degree.
Then, we can label each socket, so that, for example, the first node
in the left side of the bipartite graph would have sockets labelled
$L_{11},L_{12},\ldots L_{1\lambda_{1}}$, where the symbol $L_{ij}$
is used to denote the $j$th socket on the $i$th left node. Thus,
the $i$th left node would have $\lambda_{i}$ sockets labelled $L_{i1},L_{i2},\ldots L_{i\lambda_{i}}$.
Also, the right nodes would have sockets labelled $R_{ij}$, where
$R_{ij}$ denotes the $j$th socket on the $i$th right node. This
node and socket configuration model is a standard way to consider
the set of bipartite graphs that form the Tanner graphs of LDPC ensembles,
and in particular is discussed in length in \cite{Richardson:2008:MCT:1795974}.
A multigraph together with a labelling of the sockets of each node
is called a \emph{configuration}. Any particular left and right degree
sequences $\Lambda$ and $P$ have associated with them the set of
all configurations with these node degree sequences, and this set
is called the \emph{configuration space} associated with the degree
sequences. Clearly, a configuration is determined by a permutation
mapping the $\left|E\right|$ left node sockets to the $\left|E\right|$
right node sockets. Note that there are $\left|E\right|!$ configurations
within the space of configurations with degree sequences $\Lambda$
and $P$. Let the set of configurations with degree sequences $\Lambda$ and $P$ be denoted $\mathcal{B}\left(\Lambda, P\right)$. Since a configuration is merely a graph with a labelling of sockets for each node, graph properties can be extended to describe configurations in the natural way, including minimum bisection width.

Define
\[
B_{a}=\left\{ G\in\mathcal{B}\left(\lambda,\rho\right):\exists\mbox{ a bisection }K\in E\mbox{ such that}\left|K\right|=a\right\} 
\]
or in other words let $B_{a}$ be the set of configurations in $\mathcal{B}\left(\lambda,\rho\right)$
that have a bisection of size $a$. Note that $B_{a}$
does not represent the set of configurations in $\mathcal{B}\left(\Lambda,P\right)$
with \emph{minimum }bisection width $a$, but rather the set of graphs
with \emph{any} bisection of size $a$. Define $B_{a}^{*}$ to be
the set of all configurations in $\mathcal{B}\left(\Lambda,P\right)$ that have a bisection of size $a$ or less,
or in particular 
\[
B_{a}^{*}=\bigcup_{i=0}^{a}B_{i}.
\]

Define 
\begin{equation}
\delta_{L}\left(\Lambda\right)=\frac{1}{n}\sum_{i=\left\lfloor \frac{n}{2}\right\rfloor }^{n}\lambda_{i}\label{eq:HighestLeftDegreeSum}
\end{equation}
 (a function of a particular left degree sequence) and let 
\[
\sigma_{L}\left(\Lambda\right)=\frac{\left|E\right|}{n}-\delta_{L}.
\]
We define these quantities so that any subset of half the left nodes
can have at most $\delta n$ ``sockets'' leading from these nodes.
Similarly, define 
\[
\delta_{R}\left(P\right)=\frac{1}{m}\sum_{i=\left\lfloor \frac{m}{2}\right\rfloor }^{m}\rho_{i}
\]
 and 
\[
\sigma_{R}\left(P\right)=\frac{\left|E\right|}{m}-\delta_{R}
\]
The quantities $\delta_{L}\left(\Lambda\right)$ and $\sigma_{L}\left(\Lambda\right)$
are functions of the left degree distribution. As well, $\delta_{R}\left(P\right)$
and $\sigma_{R}\left(P\right)$ are functions of the right degree
distribution. For convenience, we may sometimes denote these quantities
as $\delta_{L}$, $\sigma_{L}$,$\delta_{R}$ and $\sigma_{R}$, and
their dependence on the degree distributions is to be implicit. Thus,
it is clear that the total number of edges in such a configuration
is $\delta_{L}n+\sigma_{L}n=\delta_{R}m+\sigma_{R}m$. Define 
\begin{equation}
\delta\left(\Lambda,P\right)=\frac{\max\left(n\delta\left(\Lambda\right),m\delta'\left(P\right)\right)}{n}\label{eq:deltaDefinition}
\end{equation}
and define
\[
\sigma\left(\Lambda,P\right)=\frac{\left|E\right|}{n}-\delta. \label{eq:sigmaDefn}
\]
For notational convenience we will abbreviate these two quantities as $\delta$ and $\sigma$ and their dependence on the node degree distribution under discussion is to be implicit. Note that $\left|E\right|=\delta n+\sigma n$. These quantities are defined
so that in any subset of half the nodes $\left(\frac{n+m}{2}\right)$
of a configuration in $\mathcal{B}\left(\Lambda,P\right)$, the minimum
of the number of left sockets and right sockets cannot exceed $\delta n$.
This observation will be useful in deriving the bounds in this and will be made more formal in Lemma \ref{lem:SubsetOfHalfNodesBound}.

Consider a given set of nodes $N\subseteq V$ for a bipartite multigraph
as defined above, with left degree sequences $\Lambda$ and right degree
sequences $P$. For a given subset of vertices $N$ we can thus divide
this set into two disjoint sets, $N_{L}$ and $N_{R}$, where $N_{L}$
is the set of all those vertices in $N$ that are left nodes, and $N_R$ all those vertices in $N$ that are right nodes.
Let $R\left(N\right)=\sum_{v\in N_{R}}\deg\left(v\right)$ and $L\left(N\right)=\sum_{v\in N_{L}}\deg\left(v\right)$
be the number of ``sockets'' attached to the left nodes in $N$
and right nodes in $N$ respectively. 
\begin{lem}
\label{lem:SubsetOfHalfNodesBound}For any bipartite multigraph $G=\left(V_{L}\amalg V_{R},E\right)$
with left degree sequences $\Lambda$ and right degree sequences $P$,
for any collection $N$ of $\frac{n+m}{2}$ vertices, $\min\left(L\left(N\right),R\left(N\right)\right)\le n\delta$.\end{lem}
\begin{rem}
We will use this lemma in a counting upper-bounding argument. Specifically,
we will count the number of graph configurations that have a bisection
of size $a$ by dividing the vertices that form a graph into two equally-sized
sets. The quantity $\min\left(L\left(N\right),R\left(N\right)\right)$
will be important for our counting bounds. \end{rem}
\begin{IEEEproof}
Suppose not. This implies that both $L\left(N\right)>n\delta$ and
$R\left(N\right)>n\delta$. Divide the vertices in $N$ into the left
nodes $N_{L}$ and right nodes $N_{R}$. It must be that $\left|N_{L}\right|+\left|N_{R}\right|=\frac{n+m}{2}$.
Thus, it must be that $\left|N_{L}\right|\le\frac{n}{2}$ or
$\left|N_{R}\right|\le\frac{m}{2}$ (otherwise their sum would exceed
$\frac{m+n}{2}$). Let us consider the case in which $\left|N_{L}\right|\le\frac{n}{2}$
(the other case leads to an analogous argument). If $\left|N_{L}\right|\le\frac{n}{2}$
and $L\left(N\right)>n\delta$, then, in particular $L\left(N\right)>n\delta_{L}\left(\Lambda\right)$
by the definition of $n\delta$. But $\delta_{L}\left(\Lambda\right)$
by definition \ref{eq:HighestLeftDegreeSum} is the sum of the highest
degree left nodes. A collection of at most half these nodes cannot
exceed this quantity, leading to a contradiction.\end{IEEEproof}
\begin{lem}
\label{lem:LemmaLabel}If a configuration $G=\left(V_{L}\amalg V_{R},E\right)$
with degree sequences $P$ and $\Lambda$ is generated according
to the uniform configuration model, then the probability that this configuration  is in the set  $B_{a}^{*}$ and hence has a bisection of
 size $a$ or less, when 
\begin{equation}
0\le a<\sigma n\label{eq:ConditionOna-1}
\end{equation}
 is upper bounded by
\end{lem}
\begin{equation}
P\left(B_{a^{*}}\right)\le\frac{\left(a+1\right)n^{2}{n \choose \frac{n}{2}}^{2}{\left|E\right| \choose a}^{4}a!\left(\delta n\right)!\left(\sigma n-a\right)!}{\left(\delta n+\sigma n\right)!}.\label{eq:LemmaOne-1}
\end{equation}

\begin{IEEEproof}
Follows from a straightforward counting upper-bounding technique given in the appendix.
\end{IEEEproof}
This lemma can be used to prove our main theorem which shows that
if a sequence of node-and-socket configurations is generated uniformly
over all such configurations, and the quantities $\delta$
and $\sigma$ (quantities that could in general change with each element of the sequence) scale according
to a particular condition, then the probability that a configuration
in this randomly generated sequence has a small bisection  (proportional to $n$ or less) approaches
$0$.

Our main theorem concerns sequences of random configurations. Specifically, we concern ourselves with a sequence of random configurations $G_1, G_2, \ldots$ where each $G_i$ in the sequence is a configuration generated according to the uniform configuration model, in which the $i$th configuration is drawn according to node degree distributions $\Lambda_i$ and $P_i$. Note that the randomness for each element of such a sequence does not come from the degree distributions: we are assuming that these distributions are fixed. It is the interconnections between nodes that is random. We specifically concern ourselves with a sequence in which the number of left nodes $n$ increases without bound. For such a sequence, denote the number of left nodes of the $i$th configuration as $n_i$. We will abbreviate the quantities $\delta\left(\Lambda_{i},P_{i}\right)$ and $\sigma\left(\Lambda_{i},P_{i}\right)$ with the symbols $\delta_i$ and $\sigma_i$ respectively, where we recall their definitions in (\ref{eq:deltaDefinition}) and (\ref{eq:sigmaDefn}). When the dependence on $i$ is clear, the subscript for these symbols may be omitted for convenience.
\begin{thm}
\label{thm:MainTheorem}Suppose that there is a sequence of randomly
generated bipartite configurations with a series of degree sequences
in which in which the number of left nodes approaches infinity, and if 
\begin{align}
\lim_{i\rightarrow\infty}2H\left(\frac{1}{2}\right)+\delta_i\left(\ln\left(\frac{\delta_i}{\delta_i+\sigma_i}\right)\right)+\nonumber \\
\sigma_i\left(\ln\left(\frac{\sigma_i}{\delta_i+\sigma_i}\right)\right) & <0\label{eq:sufficientCondition}
\end{align}
 then there exists some $\beta>0$ in which 
\[
\lim_{i\rightarrow\infty}P\left(B_{\beta n_i}^{*}\right)\rightarrow0
\]
and in particular, this occurs for any value of $0<\beta<\sigma$  that satisfies:
\begin{align}
\lim_{i\rightarrow\infty} 2H\left(\frac{1}{2}\right)+4H\left(\frac{\beta}{\delta_i+\sigma_i}\right)+\beta\left(\ln\left(\frac{\beta}{\sigma_i-\beta}\right)\right)\nonumber \\
+\delta_i\left(\ln\left(\frac{\delta_i}{\delta_i+\sigma_i}\right)\right)+\sigma_i\left(\ln\left(\frac{\sigma_i-\beta}{\delta_i+\sigma_i}\right)\right) & <0.\label{eq:conditionOnBetaWithRespectTodeltaSigma}
\end{align}
\end{thm}
\begin{rem}
This theorem says that subject to some condition on the average edge
degrees of the configurations, as these configurations get larger the probability that
the configuration generated has a bisection proportional to $n$ or less gets vanishingly
small. We will use this result to show that for capacity-approaching
LDPC degree distributions, the minimum bisection width must be large
in some sense, implying that circuit implementations of these LDPC
Tanner graphs must grow quickly as well, with high probability. The
condition in (\ref{eq:sufficientCondition}) recognizes that for a sequence
of such graphs, the quantities $\delta$ and $\sigma$ could change
with increasing $n$. If the condition is satisfied (which we will
see for capacity-approaching LDPC degree sequences it must) then with
high probability the graphs do not have a ``small'' bisection.\end{rem}
\begin{IEEEproof}
(of Theorem \ref{thm:MainTheorem}) Consider first a specific random configuration in the sequence with block length $n$ and node degree distributions that result in values for $\delta$ and $\sigma$. We will use the bounds of Lemma \ref{lem:LemmaLabel} and then apply well known approximations. Firstly, we use  the well known bounds
that
\[
e\left(\exp\left(n\ln\left(\frac{n}{e}\right)\right)\right)\le n!\le e\left(\exp\left(n\ln\left(\frac{n+1}{e}\right)\right)\right)
\]
and that
\[
{n \choose k}\le\exp\left(nH\left(\frac{k}{n}\right)\right)
\]
 where $H\left(x\right)=-x\log x-\left(1-x\right)\log\left(1-x\right)$.
We use base $e$ as opposed to base $2$ in order to conveniently
simplify the expressions that follow. Applying these bounds appropriately
to the bound in Lemma \ref{lem:LemmaLabel}, and grouping terms that grow slower than $n$ into an arbitrary polynomial term
$P\left(n\right)$ we get the following:
\begin{align*}
P\left(B^{*}_{a}\right) & \le P\left(n\right)\left(a+1\right)\exp\left(2nH\left(\frac{1}{2}\right)+4nH\left(\frac{a}{\left|E\right|}\right)\right)\\
 & \exp\left(a\ln\left(\frac{a+1}{e}\right)\right)+\delta n\ln\left(\frac{\delta n+1}{e}\right)\\
 & \exp\left(\left(\sigma n-a\right)\ln\left(\frac{\sigma n-a+1}{e}\right) \right) \\
 & \exp \left( -\left(\delta n+\sigma n\right)\ln\left(\frac{\delta n+\sigma n}{e}\right)\right).
\end{align*}
Expanding the last two terms in the exponent gives us:
\begin{align*}
P\left(B^{*}_{a}\right) & \le P\left(n\right)\left(a+1\right)\exp\left(2nH\left(\frac{1}{2}\right)+4nH\left(\frac{a}{\left|E\right|}\right)\right)\\
 & \exp\left(a\ln\left(\frac{a+1}{e}\right)\right)+\delta n\ln\left(\frac{\delta n+1}{e}\right)\\
 & \exp\left(\left(\sigma n\right)\ln\left(\frac{\sigma n-a+1}{e}\right)-a\ln\left(\frac{\sigma n-a+1}{e}\right)\right)\\
 & \exp\left(-\left(\delta n\right)\ln\left(\frac{\delta n+\sigma n}{e}\right)-\left(\sigma n\right)\ln\left(\frac{\delta n+\sigma n}{e}\right)\right).
\end{align*}
Factoring the terms in the exponent with an $a$ term, a $\delta n$
term, and a $\sigma n$ term gives us:
\begin{align*}
P\left(B^{*}_{a}\right)& \le P\left(n\right)\left(a+1\right)\exp\left(2nH\left(\frac{1}{2}\right)+4nH\left(\frac{a}{\left|E\right|}\right)\right)\\
 & \exp\left(a\left(\ln\left(\frac{a+1}{e}\right)-\ln\left(\frac{\sigma n-a+1}{e}\right)\right)\right)\\
 & \exp\left(\left(\sigma n\right)\left(\ln\left(\frac{\sigma n-a+1}{e}\right)-\ln\left(\frac{\delta n+\sigma n}{e}\right)\right)\right)\\
 & \exp\left(\left(\delta n\right)\left(\ln\left(\frac{\delta n+1}{e}\right)-\ln\left(\frac{\delta n+\sigma n}{e}\right)\right)\right).
\end{align*}
Simplifying the logarithmic expressions in each line gives us:
\begin{align*}
P\left(B^{*}_{a}\right)& \le P\left(n\right)\left(a+1\right)\exp\left(2nH\left(\frac{1}{2}\right)+4nH\left(\frac{a}{\left|E\right|}\right)\right)\\
 & \exp\left(a\left(\ln\left(\frac{a+1}{\sigma n-a+1}\right)\right)\right)\\
 & \exp\left(\left(\sigma n\right)\left(\ln\left(\frac{\sigma n-a+1}{\delta n+\sigma n}\right)\right)\right)\\
 & \exp\left(\left(\delta n\right)\left(\ln\left(\frac{\delta n+1}{\delta n+\sigma n}\right)\right)\right).
\end{align*}
We now let $a=\beta n$, which will satisfy the condition specified
in (\ref{eq:ConditionOna-1}) for $\beta<\sigma$. Making
this substitution and also using that $\left|E\right|=\delta n+\sigma n$
to expand the $\left|E\right|$ term in the first line of the expression
gives us:
\begin{align*}
& P\left(B^{*}_{\beta n}\right)  \le 
 P\left(n\right)\left(\beta n+1\right) \\
& \exp\left(2nH\left(\frac{1}{2}\right)+4nH\left(\frac{\beta n}{\delta n+\sigma n}\right)\right)\\
 & \exp\left(\beta n\left(\ln\left(\frac{\beta n+1}{\sigma n-\beta n+1}\right)\right)\right)\\
 & \exp\left(\left(\sigma n\right)\left(\ln\left(\frac{\sigma n-\beta n+1}{\delta n+\sigma n}\right)\right)\right)\\
 & \exp\left(\left(\delta n\right)\left(\ln\left(\frac{\delta n+1}{\delta n+\sigma n}\right)\right)\right).
\end{align*}
Simplifying each quotient within the logarithms, and grouping the
$\left(\beta n+1\right)$ term into our arbitrary polynomial term:
\begin{align*}
P\left(B^{*}_{\beta n}\right) & \le  P\left(n\right)\exp\left(2nH\left(\frac{1}{2}\right)+4nH\left(\frac{\beta}{\delta+\sigma}\right)\right)\\
 &   \exp\left(\beta n\left(\ln\left(\frac{\beta+\frac{1}{n}}{\sigma-\beta+\frac{1}{n}}\right)\right)\right)\\
 &   \exp\left(\left(\sigma n\right)\left(\ln\left(\frac{\sigma-\beta+\frac{1}{n}}{\delta+\sigma}\right)\right)\right)\\
 &   \exp\left(\left(\delta n\right)\left(\ln\left(\frac{\delta+\frac{1}{n}}{\delta+\sigma}\right)\right)\right).
\end{align*}
By factoring the $n$ term and by applying Lemma \ref{lem:OverallLimitLemma},
we see that the above expression will approach $0$ if 
\begin{align*} 
&\lim_{i\rightarrow\infty}2H\left(\frac{1}{2}\right)+4H\left(\frac{\beta}{\delta+\sigma}\right)+\beta\left(\ln\left(\frac{\beta+\frac{1}{n}}{\sigma-\beta+\frac{1}{n}}\right)\right)\\
&+\left(\sigma\right)\left(\ln\left(\frac{\sigma-\beta+\frac{1}{n}}{\delta+\sigma}\right)\right)+\left(\delta\right)\left(\ln\left(\frac{\delta+\frac{1}{n}}{\delta+\sigma}\right)\right)  \le  0 
\end{align*}
where we recall again that the dependence on $i$ in this expression comes from the $n$ terms and the $\delta$ and $\sigma$ terms (whose dependence on $i$ we have suppressed for notational compactness). 
This is true if
\begin{eqnarray*}
\lim_{i\rightarrow\infty}2H\left(\frac{1}{2}\right)+4H\left(\frac{\beta}{\delta+\sigma}\right)+\beta\left(\ln\left(\frac{\beta}{\sigma-\beta}\right)\right)\\
+\sigma\left(\ln\left(\frac{\sigma-\beta}{\delta+\sigma}\right)\right)+\delta\left(\ln\left(\frac{\delta}{\delta+\sigma}\right)\right) & \le & 0.
\end{eqnarray*}
Also note that this is the condition
on $\beta$ given in (\ref{eq:conditionOnBetaWithRespectTodeltaSigma}).
To derive the condition in (\ref{eq:sufficientCondition}), we find
the limit as $\beta$ approaches $0$ of this expression, and treating
the other terms as constants, giving us:
\begin{eqnarray*}
2H\left(\frac{1}{2}\right)+\sigma\left(\ln\left(\frac{\sigma}{\delta+\sigma}\right)\right)\\
+\delta\left(\ln\left(\frac{\delta}{\delta+\sigma}\right)\right) & \le & 0
\end{eqnarray*}
where we have applied the easily verifiable facts that $\lim_{x\rightarrow0}H\left(\frac{x}{c}\right)=0$
and $\lim_{x\rightarrow\infty}x\left(\ln\left(\frac{x}{\sigma-x}\right)\right)=0$
to get rid of the second and third terms in the expression. Thus,
if this condition is satisfied, by the definition of a limit, there
exists a sufficiently small $\beta$ in which $\lim_{i\rightarrow\infty}P\left(B^*_{\beta n}\right)=0$.
\end{IEEEproof}
As we are considering a sequence of configurations, we let $\omega_{i}$
be the minimum bisection width of the $ith$ configuration. This Theorem
has an obvious corollary.
\begin{cor}
\label{cor:relationshipBetweenBaStarAndMinimumBisectionWidth} If
there is a sequence of configurations as described in Theorem
\ref{thm:MainTheorem}, in which the condition in (\ref{eq:sufficientCondition})
is satisfied then $\lim_{i\rightarrow\infty}P\left(\omega_{i}\ge\beta n_{i}\right)=1$.\end{cor}
\begin{IEEEproof}
Note that the event $B^{*}_{a}$ is the event that a random configuration
has a bisection of size $a$ or less. The complement of this event
is the event that a random configuration has no bisection of size
$a$ or less, and thus equal to the event that a random configuration
has minimum bisection width greater than or equal to $a$. The corollary
flows directly from this observation.
\end{IEEEproof}

\subsection{Application to a Specific Sequence of Random Configurations}

Our result in Theorem \ref{thm:MainTheorem} can be directly applied
to the Tanner graphs of specific sequences of LDPC codes. For example,
consider a regular LDPC ensemble with variable node degree $6$ and
check node degree $3$. A randomly generated Tanner graph with this
degree distribution would have $\delta n=\sum_{\frac{n}{2}}^{n}=6\frac{n}{2}=3n$
and $\sigma n=\left|E\right|-3n=3n$. In this case we can compute
that the condition in (\ref{eq:sufficientCondition}) evaluates to: 

\begin{align*}
2H\left(\frac{1}{2}\right)+\delta\left(\ln\left(\frac{\delta}{\delta+\sigma}\right)\right)+\sigma\left(\ln\left(\frac{\sigma}{\delta+\sigma}\right)\right) & =\\
2H\left(\frac{1}{2}\right)+3\left(\ln\left(\frac{3}{3+3}\right)\right)+3\left(\ln\left(\frac{3}{3+3}\right)\right) & \approx\\
 & -2.77
\end{align*}
which we see is less than $1$. Thus, applying our theorem means that
since the condition (\ref{eq:sufficientCondition}) is satisfied,
if random Tanner graphs are generated with this degree distribution,
with probability approaching $1$ the minimum bisection width of these graphs will
be proportional to $n$.

\section{\label{sec:Almost-Sure-Bounds-on-LDPC-Circuits}Almost Sure Bounds
on Capacity Approaching LDPC Circuits }

We will use the result above to find an ``almost sure'' scaling
rule for the energy of a capacity-approaching directly-implemented
decoding scheme in which the Tanner graph of each decoder is generated
according to a uniform configuration model with a set node degree
distribution.

Consider a decoding scheme $C_{1},C_{2},\ldots$ in which each of
the decoders in the scheme are directly-implemented LDPC decoders,
as in Definition \ref{Definition:DirectlyImplementedLDPCDecoder}.
We associate a scheme with a channel that the decoders are to decode.
Let the capacity of that channel be $C$. Let the $i$th decoder have
associated block length $n_{i}$. Let the rate associated with the
$i$th decoder be $R_{i}$. Let the gap to capacity associated with
the $i$th decoder be $\eta_{i}=\frac{R_{i}}{C}$. Let the area of
the $i$th decoder be $A_{i}$, and the energy of the $i$th decoder
be $E_{i}$. Let the minimum bisection width of the Tanner graph of
the $i$th decoder be $\omega_{i}$. We consider a family of LDPC
decoding schemes in which the Tanner graph of each decoder in the
scheme is generated according to a uniform configuration model. Thus,
we say that the Tanner graph of decoder $i$ is generated uniformly
from a family $\mathcal{B}_{i}\left(\Lambda,P\right)$ of configurations.
We can thus discuss the probability of the $i$th decoder having certain
properties. In particular, in the corollary below, we will analyze
$P\left(\omega_{i}\ge\beta n_{i}\right)$, the probability that the
$i$th decoder has a Tanner graph with minimum bisection width greater
than $\beta n_{i}$, and show that this approaches $1$, resulting
in an almost sure energy scaling rule for capacity-approaching LDPC
decoders. We let the event that the $i$th decoder has a bisection
of size $a$ or less to be $B_{i,a}^{*}$
\begin{cor}
\label{cor:LDPCDirectlyScalingRule}For a family of capacity-approaching
directly-implemented LDPC decoding schemes where the Tanner graph
of each decoder is generated according to a uniform configuration
model, $\lim_{i\rightarrow\infty}P\left(A_{i}\ge cn_{i}^{2}\right)=1$
for some constant $c>0$. Similarly, $\lim_{i\rightarrow\infty}P\left(A_{i}\ge\frac{c'}{\left(1-\eta_{i}\right)^{4}}\right)=1$
for a constant $c'>0$. \end{cor}
\begin{IEEEproof}
Note that a Tanner graph is in fact a bipartite graph as described
in the Theorem \ref{thm:MainTheorem} in which the block length corresponds to $n$
and the number of checks corresponds to $m$. For a sequence of LDPC
codes to approach capacity, the result in \cite{Sason} implies that 

\[
\frac{\left|E\right|}{n\left(1-R\right)}\ge\Omega\left(\ln\left(\frac{1}{1-\eta}\right)\right)
\]
Thus, as capacity is approached, the number of edges per node must
approach infinity, and thus the quantity $\delta$ must approach
infinity. We can thus show that the expression:
\begin{equation}
2H\left(\frac{1}{2}\right)+\delta\left(\ln\left(\frac{\delta}{\delta+\sigma}\right)\right)+\sigma\left(\ln\left(\frac{\sigma}{\delta+\sigma}\right)\right)<0\label{eq:inequalityToBeSatisfied}
\end{equation}
must be satisfied for sufficient closeness to capacity.

To see this, note that $\delta$ approaches $\infty$ for a capacity-approaching code. What happens to $\sigma$ is either (a) $\lim_{n\rightarrow\infty}\frac{\delta}{\delta+\sigma}<1$
or (b) $\lim_{n\rightarrow\infty}\frac{\delta}{\delta+\sigma}=1$,
or (c) this limit does not exist. Note that this value cannot exceed $1$
because necessarily $\sigma\le\delta$. 

In the case of (c), it must be that the value
of $\sigma$ alternates and no limit can be defined. In this case,
however, we should consider the specific subsequence of decoders in
which either (a) or (b) applies. It will be clear that since for each
subsequence the appropriate scaling rule holds, thus it must be true
for the entire sequence.

In case (a): In the limit, $\ln\left(\frac{\delta}{\delta+\sigma}\right)<0$
and so $\delta\left(\ln\left(\frac{\delta}{\delta+\sigma}\right)\right)\rightarrow-\infty$,
as $\delta$ approaches $\infty$. Since $\sigma\left(\ln\left(\frac{\sigma}{\delta+\sigma}\right)\right)<0$
in any case (a consequence of $\sigma\le\delta$), thus in the limit
the inequality (\ref{eq:inequalityToBeSatisfied}) will be satisfied.

For case (b), in which $\ln\left(\frac{\sigma}{\delta+\sigma}\right)\rightarrow-\infty$,
note that $\sigma$ is positive, so $\sigma\left(\ln\left(\frac{\sigma}{\delta+\sigma}\right)\right)\rightarrow-\infty$
, and thus in the limit (\ref{eq:inequalityToBeSatisfied}) will also
be satisfied.

Note that each Tanner graph in the sequence under consideration is
generated according to the uniform configuration model. Since the sequence
is capacity approaching, by the argument above the node degree distributions
satisfy the sufficient condition of Theorem \ref{thm:MainTheorem}.
Thus, by applying Corollary \ref{cor:relationshipBetweenBaStarAndMinimumBisectionWidth},
\begin{equation}
\lim_{i\rightarrow\infty}P\left(w_{i}\ge\beta n_{i}\right)=1.\label{eq:almostSureMBWScaling}
\end{equation}

We combine this result with Thompson's \cite{Thompson} result presented
in Lemma \ref{lem:ThompsonLemma} that the area of a VLSI instantiation
of a graph with minimum bisection width $\omega$ is lower bounded by
$A_{\mathrm{c}}\ge\frac{\lambda_{\mathrm{w}}^{2}\omega^{2}}{4}.$ Thus, the event that
$\omega_{i}\ge\beta n_{i}$ implies that $A_{i}\ge\frac{\lambda_{\mathrm{w}}^{2}\left(\beta n_{i}\right)^{2}}{4}$
and thus, 

\[
\lim_{i\rightarrow\infty}P\left(A_{i}\ge\frac{\lambda_{\mathrm{w}}^{2}\left(\beta n_{i}\right)^{2}}{4}\right)=1
\]
as expressed in the theorem statement. 

This result can be used to understand how the area of almost all circuits
that instantiate random Tanner graphs of LDPC codes must scale as
capacity is approached. It is well known from \cite{GallagerBook, Strassen}
that, as a function of fraction of capacity $\eta=\frac{R}{C}$, the
minimum block length required for any code scales as:

\[
n\approx\frac{b}{\left(1-\eta\right)^{2}}
\]
for a constant $b$ that depends on the channel statistics and also
the target probabilities of error. We are not concerned with the value
of this constant but rather the dependence of this expression on $\eta$.

We use this to note that, if $\omega_{i}\ge\beta n_{i}$, then, recognizing
from Definition \ref{Definition:DirectlyImplementedLDPCDecoder} that
a directly-instantiated LDPC decoder must contain its Tanner graph,
and also applying Lemma \ref{lem:AMinLemma} which says that a circuit
must be bigger than the minimum area of a circuit instantiation of
a graph that the circuit contains, then 
\[
A_{\mathrm{c}}\ge\frac{\lambda_{\mathrm{w}}^{2}\beta^{2}n^{2}}{4}\ge\frac{\lambda_{\mathrm{w}}^{2}\beta^{2}}{4}\frac{b^{2}}{\left(1-\eta\right)^{4}}\ge\Omega\left(\frac{1}{\left(1-\eta\right)^{4}}\right).
\]
Combining this observation with the result in (\ref{eq:almostSureMBWScaling})
results in 
\[
\lim_{i\rightarrow\infty}P\left(A_{i}\ge\frac{c'}{\left(1-\eta_{i}\right)^{4}}\right)=1
\]
 for a constant $c'>0$, finishing the proof.
\end{IEEEproof}

\subsection*{Applicability of this Result}

There is a minor detail that needs to be dealt with for this theorem
to be truly useful. Our results assume that a Tanner graph is directly
implemented in wires. This is indeed a practical way to create a decoding
circuit. However, according to our configuration model, it is possible
that two or more edges can be drawn between the same two nodes. This
type of conflict is usually dealt with by deleting even multi-edges
and replacing odd multi-edges with a single edge (see definition
3.15, the Standard LDPC Ensemble in \cite{Richardson:2008:MCT:1795974}).
This leads to a potential problem with the applicability of our theorem:
what happens if the edges that we delete form a minimum bisection
of the induced graph? In that case it is possible that the graph we
instantiate on the circuit has a lower minimum bisection width than
that which we calculated, and thus could possibly have less area.
However, this is resolved by the fact that in the limit as $n$ approaches
infinity for a standard LDPC ensemble, the graph is locally tree-like
(Theorem 3.49 in \cite{Richardson:2008:MCT:1795974}) with probability
approaching $1$. This implies that the probability that the number
of multi-edges in a randomly generated configuration is some fraction
of $n$ must approach $0$ (or else the graph would not be locally
tree-like, contradicting the theorem). Hence, even if we did delete
these multi-edges from the randomly generated configuration, this
could at most decrease the minimum bisection width by the number of
deletions, but this number of deletions, with probability $1$, cannot
grow linearly with $n$. Hence, the minimum bisection width must still,
with probability $1$, grow linearly with $n$, and our scaling rules
are still applicable.

\subsection{Energy Complexity of Capacity Approaching Complete-Check-Node LDPC Decoders}

Below we will consider a sequence of capacity-approaching, complete-check-node
serialized decoders. Recall that these decoders do not directly instantiate their Tanner graph in
wires, but they do have subcircuits corresponding to each check and variable node. In each iteration,
possibly over several clock cycles, messages are to be passed from each variable node subcircuit to their corresponding
check node subcircuit and similarly for the check node subcircuits passing messages to their corresponding variable node subcircuits. It may be that the same wire is used to transmit different messages during different clock cycles of the same iteration
of the computation. It is thus possible that such a method can decrease wiring area (by not requiring a wire for each edge of the
Tanner graph) at the cost of more clock cycles. We prove below that such a method still results in a super-linear almost sure
lower bound on energy complexity. So there is no ambiguity, a sequence of decoders for a channel with capacity $C$ with rates
$R_1, R_2, \ldots$ is capacity-approaching if $\lim_{i\rightarrow\infty}R_{i}=C$.

\begin{cor} For a sequence of capacity-approaching, complete-check-node
serialized LDPC decoders whose Tanner graphs are generated according
to the uniform configuration model, $\lim_{i\rightarrow\infty}P\left(E_{i}\ge cn_{i}^{1.5}\right)=1$
for some $c>0$. Also, $\lim_{i\rightarrow\infty}P\left(E_{i}\ge\frac{c}{\left(1-\eta\right)^{3}}\right)=1$ and $\lim_{i\rightarrow\infty}P\left(\frac{E_{i}}{k}\ge\frac{c}{1-\eta}\right)=1$.
\end{cor}
\begin{IEEEproof}
In considering a complete-check-node serialized LDPC decoder, we
note that such a decoder contains a graph with $n$ variable nodes
and at least $n-k$ check nodes. We will use arguments similar to those used
by Thompson \cite{Thompson} and Grover \cite{groverFundamental}. Let the minimum bisection width of the Tanner graph
of the associated with the $i$th decoder be $\omega_{i}$. Suppose
that the graph of the circuit implementing this decoder has minimum
bisection width $W_{i}$ (we use the symbol $W_i$ to distinguish this
from the minimum bisection width of the Tanner graph $\omega_{i}$
of the $i$th decoder, recalling that we do not require in this case that the circuit contains the underlying Tanner graph). Thus, in one iteration, the number of bits
communicated between any bisection of the nodes must at least be $\omega_{i}$.
One iteration must be performed, but since the minimum bisection width
of the graph associated with this circuit is $W_{i}$, this requires
that more clock cycles are used to pass the information between the
check and variable nodes, and in particular
\begin{equation}
\tau_{i}W_{i}\ge\omega_{i}.\label{eq:simpleTauOmegaBoundSerialized}
\end{equation}
We also know from Lemma \ref{lem:ThompsonLemma} that $A_{i}\ge\frac{\lambda_{\mathrm{w}}^{2}W_{i}^{2}}{4}$
and so combining with the inequality in (\ref{eq:simpleTauOmegaBoundSerialized})
gives us:

\begin{equation}
A_{i}\tau^{2}_{i}\ge\frac{\lambda_{\mathrm{w}}^{2}W_{i}^{2}\tau_{i}^{2}}{4}\ge\frac{\lambda_{\mathrm{w}}^{2}\omega_{i}^{2}}{4}.\label{eq:AcTauSquaredBound}
\end{equation}
Trivially, because there are $n_i$ variable node subcircuits in the circuit, $A_i\ge n_i$
and thus combining with (\ref{eq:AcTauSquaredBound}) we get
\[
A_{i}^{2}\tau_{i}^{2}\ge\frac{\lambda_{\mathrm{w}}^{2}\omega_{i}^{2}n_i}{4}
\]
and thus, taking the square root of both sides of this inequality,
\[
A_{i}\tau_{i}\ge\frac{\lambda_{\mathrm{w}}\omega_{i}n^{0.5}_{i}}{2}.
\]
Since energy is proportional to the product of circuit area and number
of clock cycles, this implies that for each decoder in the sequence
\[
E_{\mathrm{i}}\ge \xi_{\mathrm{tech}}\omega_{i}n^{0.5}_{i}.
\]
for the constant $\xi_{\mathrm{tech}}$ that depends on the specific technology used to implement the circuits.

Using the same arguments as Corollary \ref{cor:LDPCDirectlyScalingRule} we can show that for a
capacity-approaching LDPC scheme $\lim_{i\rightarrow\infty}P\left(\omega_{i}\ge\beta n_{i}\right)=1$
for some $\beta>0$. Following the logic above, the event that $\omega_{i}\ge\beta n_{i}$
implies $E_{i}\ge\xi_{\mathrm{tech}}\beta n_{i}^{1.5}$ which thus implies $\lim_{i\rightarrow\infty}P\left(E_{i}\ge cn_{i}^{1.5}\right)=1$ for some constant $c>0$,
Also, following the same logic as in Corollary \ref{cor:LDPCDirectlyScalingRule}, $\lim_{i\rightarrow\infty}P\left(E_{i}\ge\frac{c}{\left(1-\eta\right)^{3}}\right)=1$
and $\lim_{i\rightarrow\infty}P\left(\frac{E_{i}}{k}\ge\frac{c}{1-\eta}\right)=1$.
\end{IEEEproof}

\subsection{Limitations of Result}

A goal of this research is to find fundamental
bounds on the ``energy complexity'' of capacity-approaching decoders
as a function of $\eta=\frac{R}{C}$. The result presented here does
not quite do this, but it does advise engineering by suggesting
that if $n$ is very large, one can be reasonably sure that the area
of a circuit that instantiates a randomly generated Tanner graph will have area that scales
as $\Omega\left(n^{2}\right)$. Of course, we have assumed that this
Tanner graph has been generated by going to each socket of the left nodes
and randomly finding a connection to a remaining right socket. This
is of course a very natural way to generate Tanner graph, and is in
fact used in the analysis of LDPC codes \cite{Richardson:2008:MCT:1795974}.

This is not to say, of course, that there don't exist good LDPC coding schemes
with slower scaling laws. Creating a sequence of LDPC
codes that avoids this scaling law with probability greater than $0$
would be possible if the random generation rule for the LDPC graph
was somehow altered. For example, perhaps the variable nodes and check
nodes could be placed uniformly scattered through a grid and then
the randomly placed edges, instead of being chosen uniformly over
all possible edges, are chosen uniformly over a choice of edges connecting
variable and check nodes that are ``close'' to each other.

In practice, a Tanner graph is often modified to prevent interconnections that are "too far" between check and variable nodes that result in long wire length and thus higher energy  \cite{RothEtAl}. Simulation in a particular case can analyze whether this technique is worth the possible code performance trade-off. Currently, however, the common technique of generating an LDPC ensemble and analyzing average code performance does not consider energy complexity as a fundamental parameter to be traded-off with other code parameters. It seems likely that if ``neighbors'' of a variable node are restricted to those check nodes that are spatially close by, an LDPC code could still have good asymptotic performance if block lengths grow large. An analysis challenge of such a scheme may be to show that asymptotically a Tanner graph generated from such a distribution is locally tree-like. Furthermore, analysis of the required block length using such a technique to get good performance would be needed: even if asymptotically such schemes perform well, it may be that much longer block lengths are required for the same performance. The cost of possibly larger block length for such a scheme would have to be considered to determine whether it is worth it to have a slower scaling rule as a function of block length if it comes at a cost of much longer block length.

Whether or not such a sequence of LDPC codes would give good performance
is unclear. However, in the
following section we can use known bounds on the average node degree
of an LDPC decoder as well as bounds on the area of graphs instantiated
on a circuit to get scaling rules that are true for \emph{all } directly-implemented capacity-approaching LDPC decoders, not just almost
all.

\section{Bounds for All LDPC Decoder Circuits\label{sec:Bounds-for-All}}

We can find bounds for the energy complexity for all capacity-approaching directly-implemented LDPC codes (and not just almost all) by using the following
Theorem:
\begin{thm}
\label{thm:CircuitEdgeNodeBound}If a circuit contains a graph $G=\left(V,E\right)$
that has no loops, according to the standard VLSI model, the total
area of a circuit that contains that graph is bounded as:
\[
A\ge\frac{\lambda_{\mathrm{w}}^{2}\left(\sqrt{2}-1\right)^{2}}{4}\frac{\left|E\right|^{2}}{\left|V\right|}
\]
where we recall that $\lambda_{\mathrm{w}}$ is the wire width in
the circuit, and $\left|E\right|$ and $\left|V\right|$ are
the number of edges and vertices in the graph, respectively.\end{thm}

The proof of this theorem uses a similar approach as used by Grover \emph{et
al.} in \cite{groverFundamental}, in which the $A_{\mathrm{c}}\tau$
complexity of circuits is related to the bits communicated within the circuit.
The result of this paper, however, is a bound on the area of a circuit instantiation
of a graph as a function of the number of edges and vertices in the graph. We use a similar nested bisection technique as the Grover \emph{et	al.} paper. The proof is given in the appendix.

This result, combined with the results in \cite{Sason} on the average
edge degree as a function of gap to capacity, results in the following
corollary:
\begin{cor}
\label{cor:LDPCSureBound}The energy of any directly-instantiated
LDPC decoder must have asymptotic energy that is lower bounded by:
\[
E_{\mathrm{dec}}\ge\Omega\left(\frac{N}{\left(1-\eta\right)^{2}}\ln^{2}\left(\frac{1}{1-\eta}\right)\right)
\]

and average energy per bit decoded that scales as

\[
\frac{E_{\mathrm{dec}}}{k}\ge\Omega\left(N\ln^{2}\left(\frac{1}{1-\eta}\right)\right)
\]
where $N$ is the number of iterations required to decode.\end{cor}
\begin{rem}
Note that the number of iterations $N$ in the above Corollary in
general may be a function of the particular decoding algorithm instantiated
and possibly the particular received vector. Our discussion does not
analyze the number of iterations required, so we simply write our scaling rules in terms
of this quantity. \end{rem}
\begin{IEEEproof}
We can combine Sason's \cite{Sason} result that the average parity node degree of the Tanner graph
of a capacity-approaching LDPC code must scale as $\Omega\left(\ln\left(\frac{1}{1-\eta}\right)\right)$
and that the minimum block length of any code must scale as $\Omega\left(\frac{1}{\left(1-\eta\right)^{2}}\right)$ \cite{GallagerBook,Strassen},
meaning that $\left|E\right|\ge\Omega\left[\left(n-k\right)\ln\left(\frac{1}{1-\eta}\right)\right]$.
Note also that the number of nodes in this graph must be at least $\left|V\right|=2n-k=O\left(n\right)$.
Combining these results along with Theorem \ref{thm:CircuitEdgeNodeBound}
results in the scaling laws in the corollary.
\end{IEEEproof}
We note that this lower bound on directly-implemented Tanner graphs
contrasts with the lower bounds in \cite{BlakeKschischangFundamentalLowerBoundArxiv}, which show an $\Omega\left(\left(\ln\left(\frac{1}{1-\eta}\right)\right)^{\frac{1}{2}}\right)$
lower bound for the per bit energy complexity of fully-parallel decoding
algorithms as a function of gap to capacity. This result means that
directly-instantiated LDPC decoders are \emph{necessarily }asymptotically
worse than this lower bound (albeit a lot closer than the $\Omega\left(\frac{1}{\left(1-\eta\right)^{2}}\right)$
almost sure lower bound of Corollary \ref{cor:LDPCDirectlyScalingRule}).
Of course, it is not known whether the lower bounds of the paper in
\cite{BlakeKschischangFundamentalLowerBoundArxiv} are tight, but Corollary \ref{cor:LDPCSureBound} proves
that directly instantiated LDPC decoders cannot reach these lower
bounds in an asymptotic sense.

\section{Conclusion\label{sec:Conclusion}}

\begin{table*}
\centering
\begin{tabular}{|c|c|}
\hline 
 & Lower Bound Scaling Rule Per Bit ($\frac{E_{\mathrm{dec}}}{k}$)\tabularnewline
\hline 
\hline 
Almost all directly instantiated LDPC decoders & $\Omega\left(N\ln^{2}\left(\frac{1}{1-\eta}\right)\right)$\tabularnewline
\hline 
Almost all LDPC decoders & $\Omega\left(\frac{N}{\left(1-\eta\right)^{2}}\right)$\tabularnewline
\hline 
All LDPC with Tanner Graph Directly Implemented & $\Omega\left(N\ln^{2}\left(\frac{1}{1-\eta}\right)\right)$\tabularnewline
\hline 
All Fully-Parallel Decoders \cite{BlakeKschischangFundamentalLowerBoundArxiv} & $\Omega\left(\sqrt{\ln\left(\frac{1}{1-\eta}\right)}\right)$\tabularnewline
\hline 
\end{tabular}
\caption{\label{tab:SummaryOfResults}Summary of the scaling rule lower bounds
derived in this paper. We present these bounds as a function of $\eta=\frac{R}{C}$.
In the first three scaling rules presented, $N$ is the number of
iterations required (which in general may be a function of the actual
LDPC code instantiated, as well as the particular received vector).
For comparison, we also include a result on lower bounds for all fully-parallel decoders
given in \cite{BlakeKschischangFundamentalLowerBoundArxiv}.}

\end{table*}

The main contribution of this paper is graph theoretic in nature.
We have shown that subject to a mild condition on node degree distributions, almost all Tanner graph instantiations
have a minimum bisection width that scales as $\Omega\left(n\right)$
where $n$ is the number of left nodes. The minimum bisection width of
a graph is related to the area of circuit implementations of these
graphs. We have used this result to show that almost all LDPC decoders
that directly instantiate their Tanner graph must have
circuit area, and thus energy, that scales as $\Omega\left(n^{2}\right)$. We can use
this result to provide a scaling rule for the energy complexity of
almost all capacity-approaching LDPC decoders. We have further presented
a general theorem on the area of circuits that instantiate any graph
to further bound the area of any LDPC decoder that approaches capacity.
These results are summarized in Table \ref{tab:SummaryOfResults}. Note that
our results show that directly-instantiated LDPC codes cannot reach the lower bounds
presented in \cite{BlakeKschischangFundamentalLowerBoundArxiv}, thus indicated that either the lower bound cited is not tight, or directly-instantiated LDPC codes asymptotically not optimal from this energy perspective. It may also be that both are true, namely that known lower bounds are not tight and LDPC codes are not asymptotically optimal. This remains an open question.

\appendices{}

\renewcommand{\thesubsection}{\Alph{subsection}}

\section{Proof Of Lemma~\ref{lem:LemmaLabel}}\label{app:LemmaProof}

\begin{IEEEproof} (of Lemma \ref{lem:LemmaLabel})
\begin{figure} \centering 
\includegraphics[width=1.8 in]{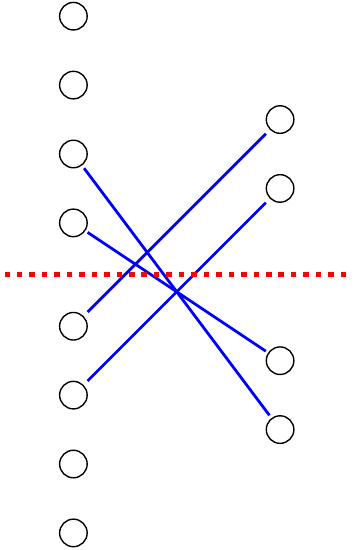}~\includegraphics[width=1.8 in]{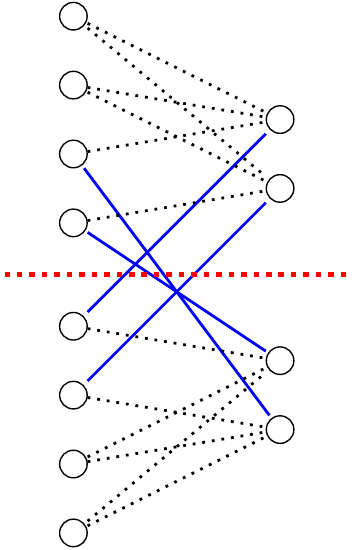}
\caption{An example of a quadrant configuration associated with a degree distribution
where all the left nodes have degree $2$ and all the right nodes
have degree $4$, in which the number of left nodes $n=8$ and number
of right nodes $m=4$. The fully drawn configuration on the right
is a quadrant configuration in $Q_{4}^{4,2}$. Recall that the superscript
denotes that there are $i=4$ top left nodes and $j=2$ edges leading
from top left nodes to bottom right nodes. The subscript indicates
that there are $a=4$ edges between top and bottom nodes, and in this
case we see that they cross a dotted line, indicating where the bisection
occurs. The diagram on the left shows the drawing of $a=4$ edges
crossing between top and bottom nodes. The graph on the right shows
a permutation of the remaining sockets in both the top and bottom
nodes.
}
\label{countingDiagram}  \end{figure}
Let the set of graphs in $\mathcal{B}\left(\Lambda,P\right)$ having
a bisection of size $a$ be denoted by $B_{a}$. Then we can say that,
according to the uniform configuration model,
the probability of the event of generating a configuration with a
bisection of size $a$ is given by: 
\[
P\left(B_{a}\right)=\frac{\left|B_{a}\right|}{\left|E\right|!}
\]
 \emph{i.e.,} it is the cardinality of the set of such configurations
divided by the total number of configurations in with node degrees $\Lambda$ and $P$.

We will now bound the number of configurations in $\mathcal{B}\left(\Lambda,P\right)$
with a bisection of size $a$, and we will assume that $a<\sigma n$.
To do so, we will define a \emph{quadrant configuration}, show that
the number of quadrant configurations with a bisection of size $a$
is greater than or equal to $B_{a}$, and then upper bound the number
of quadrant configurations with a bisection of size $a$ or less.

A quadrant configuration of a bipartite configuration $G=\left(V_{L}\amalg V_{R},E\right)$
is an ordered-tuple $Q=\left(G,T_{L},T_{R},B_{L},B_{R}\right)$ where
the vertices are divided into $4$ disjoint sets, the \emph{top left
vertices } (\emph{$T_{L}$}), the \emph{top right vertices }(\emph{$T_{R}$}),
the \emph{bottom left vertices} ($B_{L}$), and the \emph{bottom right
vertices } ($B_{R}$)\emph{,} in which $T_{L},B_{L}\subseteq V_{L}$,
$T_{R},B_{R}\subseteq V_{R}$ and $\left|\left|T_{R}\cup T_{L}\right|-\left|B_{L}\cup B_{R}\right|\right|\le1$.
Naturally, vertices in $T_{L}$ are considered \emph{top left vertices},
or, interchangeably, \emph{top left nodes,} and similarly for the
other sets of vertices in a quadrant configuration. Furthermore, vertices
in $T_{L}$ and $T_{R}$ are considered to be \emph{top vertices }or
\emph{top nodes, }and similarly for the bottom vertices.

Note that every bipartite graph has at least one quadrant configuration
induced by arbitrarily dividing the vertices in half, and denoting
one half of these vertices top vertices and the other half bottom
vertices. Thus, the set of quadrant configurations with a particular
degree distribution is at least as big as the set of configurations
with a particular degree distribution. Because a quadrant configuration
$Q=\left(G,T_{L},T_{R},B_{L},B_{R}\right)$ contains a graph $G$,
graph properties can be extended to describe a quadrant configuration.
So, for example, if we say that a quadrant configuration has minimum
bisection width $a,$ we mean precisely that the graph $G$ within
the quadrant configuration has minimum bisection width $a$.

Denote the set of quadrant configurations with set node degree distributions
$\Lambda$ and $P$ in which $a$ is the number of edges leading from top vertices to bottom vertices as $Q_{a}$. Note that
the dependence of $Q_{a}$ on a particular node degree distribution
is implicit. Observe that every configuration with a bisection of
size $a$ has a corresponding quadrant configuration in $Q_{a}$ created
in the natural way by denoting one bisected set of vertices as the
top vertices, and the other the bottom vertices. Thus $\left|B_{a}\right|\le\left|Q_{a}\right|$.

For ease of discussion, we will assume that the total number of nodes
$m+n$ in the set of configurations under discussion is even, so that
$\frac{m+n}{2}$ is an integer.

Denote the set of quadrant configurations with a bisection of size
$a$ in which there are $i$ top left nodes and $j$ edges connecting
top left vertices to the bottom right by $Q_{a}^{i,j}$. This of course
implies that there are $\frac{m+n}{2}-i$ top right nodes and $a-j$
edges leading from the bottom left to the top right nodes. We can
see in Figure \ref{countingDiagram} an example of such an element
that we are counting for the case of $n=8$ and $a=4$, $i=4$ and
$j=2$. Note then that 
\[
Q_{a}=\bigcup_{i=0}^{n}\bigcup_{j=0}^{a}Q_{a}^{i,j}
\]
We bound the size of $Q_{a}^{i,j}$ by counting all quadrant configurations
with a bisection of size $a$ that are the edges connecting top nodes
to bottom nodes. We have
\begin{align}
\left|Q_{a}^{i,j}\right| & \le\underset{a}{\underbrace{{n \choose i}}}\underset{b}{\underbrace{{m \choose \frac{m+n}{2}-i}}}\underset{c}{\underbrace{{\left|E\right| \choose j}{\left|E\right| \choose j}{\left|E\right| \choose a-j}{\left|E\right| \choose a-j}}}\nonumber \\
 & \underset{d}{\underbrace{\left(j\right)!\left(a-j\right)!}}\underset{e}{\underbrace{\left(\delta n\right)!\left(\sigma n-a\right)!}},\label{eq:BoundingBai}
\end{align}
where

a. Represents a choice of $i$ top left nodes.

b. Represents a choice of $\frac{m+n}{2}-i$ top right nodes from
the $m$ total right nodes.

c. The quantity ${\left|E\right| \choose j}$ is an upper bound on
the number of choices of $j$ sockets that will have edges that cross
the bisection line chosen from the top variable nodes, and ${\left|E\right| \choose j}$
is an upper bound on the number of choices for the bottom right sockets
to which these edges will be connected. For a configuration in $B_{a}^{i,j}$
there must also be $a-j$ edges leading from the bottom left to the
top right. The quantity ${\left|E\right| \choose a-j}$ is an upper
bound on the number of choices of sockets in the bottom left that
can have edges crossing the middle bisection, and similarly ${\left|E\right| \choose a-j}$
is an upper bound on the number of choices for the sockets connected
in the top right.

d. Counts the number of permutations of edges that join the top half
to the bottom half (first counting the $j$ connections from the top
left nodes to the bottom right nodes, then the $a-j$ connections
from the bottom variable nodes to the top variable nodes.

e. This step in the quadrant configuration construction process involves
permuting the connections of the remaining sockets in the top half
and the bottom half. However, at this point it is not clear how many
sockets are in the top half or the bottom half. However, we can upper
bound the number of permutations possible. The number of nodes available
in the top left vertices must equal the number of nodes available
in the top right vertices (because in order to construct a valid configuration
this must be true). By construction, the total number of nodes in
the top left and top right is $\frac{m+n}{2}$, and thus the number
of sockets available cannot exceed $\delta n$, by Lemma \ref{lem:SubsetOfHalfNodesBound}.
Suppose the number of sockets available for all the top left nodes
is $M$ and the sockets available in the bottom left nodes is $N$.
Then there are at most $M!N!$ ways to permute these. We also know
that $M+N=\left|E\right|-a$ (since the total number of sockets available
on one side of the constructed quadrant configuration is $\left|E\right|$ and
$a$ have been used to cross between top nodes and bottom nodes),
and that $M\le\delta n$ and $N\le\delta n$. Subject to these restrictions,
a direct application of Lemma \ref{lem:MFactorialNFactorialUpperBoundLemma} implies
$M!N!\le\left(\delta n\right)!\left(\left|E\right|-\delta n-a\right)!=\left(\delta n\right)!\left(\sigma n-a\right)!$

Now, for the sake of simplicity, we will further loosen these bounds
by upper bounding each of the factors a, b, c, and d. Each of these
bounds is easily verified:

a. We note that ${n \choose i}\le{n \choose \frac{n}{2}}$. 

b. Since $m\le n$, thus ${m \choose \frac{m+2}{2}-i}\le{n \choose \frac{n}{2}}$. 

c. ${\left|E\right| \choose j}{\left|E\right| \choose a-j}{\left|E\right| \choose j}{\left|E\right| \choose a-j}\le{\left|E\right| \choose a}^{4}$
which is implied by $a\le \sigma n \le \frac{\left|E\right|}{2}$.

d. $\left(j\right)!\left(a-j\right)!\le a!$ which flows directly
from the observation that ${a \choose j}\ge1$.

Combining these gives us the following bound:
\[
\left|Q_{a}^{i,j}\right|\le{n \choose \frac{n}{2}}^{2}{\left|E\right| \choose a}^{4}a!\left(\delta n\right)!\left(\sigma n-a\right)!
\]
We can bound $\left|Q_{a}\right|$ by summing over our upper bound
on $\left|Q_{a}^{i,j}\right|$:
\begin{eqnarray}
\left|Q_{a}\right| & \le & \sum_{i=1}^{n}\sum_{j=1}^{n}\left|Q_{a}^{i,j}\right|\nonumber \\
 & \le & n^{2}{n \choose \frac{n}{2}}^{2}{\left|E\right| \choose a}^{4}a!\left(\delta n\right)!\left(\sigma n-a\right)!\label{eq:BoundWeWillUse}
\end{eqnarray}
We of course are not concerned with the probability of a bisection
of size $a$, but rather with the probability of a bisection of size
$a$ or less. We denote the set of configurations with a bisection
of size $a$ or less by $Q_{a}^{*}$ and since $Q_{a}^{*}=\bigcup_{i=0}^{a}Q_{a}$:
\[
\left|Q_{a}^{*}\right|\le\sum_{i=0}^{a}\left|Q_{i}\right|.
\]
We will now show that the expression in (\ref{eq:BoundWeWillUse}) is
an non-decreasing function of $a$ for $0<a\le\frac{\left|E\right|-1}{2}$.
Let the right side of the expression be denoted $d_{a}$, then it
is easy to show that $\frac{d_{a+1}}{d_{a}}$ is greater than or equal
to $1$. It is easy to show that
\[
\frac{d_{a+1}}{d_{a}}=\frac{{\left|E\right| \choose a+1}^{4}\left(a+1\right)}{{\left|E\right| \choose a}^{4}\left(\sigma n-a\right)}.
\]
Expanding the binomial coefficients in the numerator and denominator and
simplifying gives us
\[
\frac{d_{a+1}}{d_{a}}=\frac{\left(\left|E\right|-a\right)^{4}}{\left(a+1\right)^{3}\left(\sigma n-a\right)}
\]
This quantity will be greater than or equal $1$ if $\left|E\right|-a\ge a+1$
and $\left|E\right|-a\ge\sigma n-a$ . Note that $a<\sigma n$ (an assumption of our lemma) implies $2a<2\sigma n\le\left|E\right|$.
Since $a$ and $\left|E\right|$ are both integers, this implies $2a\le\left|E\right|-1$,
from which we can see that the first inequality is satisfied. The second is satisfied by the fact that $\sigma n\le\left|E\right|$.
We thus observe that,
\begin{align}
\left|B_{a}^{*}\right|  &\le  \left|Q_{a}^{*}\right| \nonumber \\ &\le \sum_{i=0}^{a}\left|Q_{i}\right|\nonumber \\
  &\le  \sum_{i=0}^{a}n^{2}{n \choose \frac{n}{2}}^{2}{\left|E\right| \choose a}^{4}a!\left(\delta n\right)!\left(\sigma n-a\right)!\nonumber \\
  &\le  \left(a+1\right)n^{2}{n \choose \frac{n}{2}}^{2}{\left|E\right| \choose a}^{4}a!\left(\delta n\right)!\left(\sigma n-a\right)!\label{eq:BastarBound}
\end{align}
We note that the number of possible multi-graphs with our given node degree
distribution is at least $\left(\delta n+\sigma n\right)!$. We can
now bound the probability of the event $B^{*}_{a}$ with:
\begin{align}
P\left(B^{*}_{a}\right)                &\le\frac{\left|B^{*}_{a}\right|}{\left(\delta n+\sigma n\right)!}\\
 & \le  \frac{\left(a+1\right)n^{2}{n \choose \frac{n}{2}}^{2}{\left|E\right| \choose a}^{4}a!\left(\delta n\right)!\left(\sigma n-a\right)!}{\left(\delta n+\sigma n\right)!}
\end{align}
where we have simply applied the upper bound for the
size of $B^*_{a}$ of (\ref{eq:BastarBound}) .
\end{IEEEproof}

\section{Proof of Theorem \ref{thm:CircuitEdgeNodeBound}}\label{app:CircuitNodeEdgeTheorem}

In this section we will prove Theorem \ref{thm:CircuitEdgeNodeBound},
which states that if a circuit implements a graph $G=\left(V,E\right)$
that has no loops, according to the standard VLSI model, the total
area of that circuit is bounded by:
\[
A_{\mathrm{c}}\ge\frac{\lambda_{\mathrm{w}}^{2}\left(\sqrt{2}-1\right)^{2}}{4}\frac{\left|E\right|^{2}}{\left|V\right|}
\]
where $\lambda_{\mathrm{w}}$ is the wire width in the circuit, and $\left|E\right|$
and $\left|V\right|$ are the number of edges and vertices in the
graph, respectively.
\begin{IEEEproof}
(Of Theorem \ref{thm:CircuitEdgeNodeBound}) For simplicity we will
say the graph has $\left|V\right|=2^{k}$ vertices. Recall that a
bisection of a graph is the set of edges of that graph that divides
the vertices in half. A minimum bisection of a graph is a bisection
that uses the smallest number of edges to bisect the graph. We will perform what we call
nested minimum bisections on the graph. To do this, first, the
edges in a minimum bisection of the graph are removed, and there are
$b_{1,1}$ such edges. This divides the graph into two distinct components.
Then, these two components (which are subgraphs of the original graph)
are bisected by removing edges in their respective minimum bisection
cut, and so $b_{2,1}$ and $b_{2,2}$ edges are removed. This process
continues for $k$ bisections, and after the $k$th bisection, there
are $2^{k}$ disjoint subgraphs, each with one vertex, and no edges
(because we assume that in these graphs there are no loops). It must
be that the total of all the edges we removed equals the total number
of edges in our graph; in other words, it must be that:
\begin{equation}
\sum_{i=1}^{k}\sum_{j=1}^{2^{i-1}}b_{i,j}=\left|E\right|.\label{eq:ConstraintOnThebijs}
\end{equation}
Recall Thompson's bound from Lemma \ref{lem:ThompsonLemma} that says
for a circuit implementation of a graph with minimum bisection width
$\omega$, the area of that circuit is lower bounded by:
\[
\frac{4A_{\mathrm{c}}}{\lambda_{\mathrm{w}}}\ge\omega^{2}.
\]
We can use this result to bound the total area of all of the subgraphs
for each level $i=1,2,\ldots,k$,
\[
\frac{4A_{\mathrm{c}}}{\lambda_{\mathrm{w}}}\ge\sum_{j=1}^{2^{i-1}}b_{i,j}^{2}.
\]
Thus,
\[
\frac{4A_{\mathrm{c}}}{\lambda_{\mathrm{w}}}\ge\max\left(b_{1,1}^{2},b_{2,1}^{2}+b_{2,2}^{2},\ldots,\sum_{j=1}^{2^{k-1}}b_{k,j}^{2}\right).
\]
Standard convex optimization techniques imply that
this expression is minimized when:
\begin{align}
c_{1} & \equiv b_{1,1}\nonumber \\
c_{2} & \equiv b_{2,1}=b_{2,2}\nonumber \\
 & \vdots\nonumber \\
c_{k} & \equiv b_{k,1}=b_{k,2}=\ldots=b_{k,2^{k-1}}\label{eq:cDef-1}
\end{align}
where for the sake of convenience we have introduced the constants
$c_{1},c_{2},\ldots,c_{k}$. Furthermore, it can be shown that
\begin{equation}
b_{1,1}^{2}=b_{2,1}^{2}+b_{2,2}^{2}=\ldots=\sum_{i=1}^{2^{k-1}}b_{k,i}^{2}=a \label{eq:compSlacknessResult}
\end{equation}
for some $a$. Using the definitions of the constants $c_{1},c_{2},\ldots,c_{k}$
given in (\ref{eq:cDef-1}), and applying this to the above equation
(\ref{eq:compSlacknessResult})
\[
c_{1}^{2}=2c_{2}^{2}=4c_{3}^{2}=\ldots=2^{k-1}c_{k}
\]
from which we can infer
\[
c_{2}=\frac{1}{\sqrt{2}}c_{1},~
c_{3}=\frac{1}{\sqrt{2}}c_{2}=\left(\frac{1}{\sqrt{2}}\right)^{2}c_{1},
\]
and, in general,
\[
c_{i}=\left(\frac{1}{\sqrt{2}}\right)^{i-1}c_{1}.
\]
We then apply this to the constraint in (\ref{eq:ConstraintOnThebijs}) to give us
\begin{align*}
|E| &=
\sum_{i=1}^{k}2^{i-1}\left(\frac{1}{\sqrt{2}}\right)^{i-1}c_{1}\\
&= c_1 \sum_{i=1}^{k}\left(\frac{2}{\sqrt{2}}\right)^{i-1} \\
&= c_{1}\left(\frac{\sqrt{2}^{k}-1}{\sqrt{2}-1}\right),
\end{align*}
from which $c_1$ is easily obtained.
Now, using that $k=\log_{2}V$, we have $\sqrt{2}^{k}=\left(2^{\frac{1}{2}}\right)^{\log_{2}\left|V\right|}=\left(2^{\log_{2}\left|V\right|}\right)^{\frac{1}{2}}=\left|V\right|^{\frac{1}{2}}$.
Hence, we conclude that
\[
\frac{4A_{\mathrm{c}}}{\lambda_{\mathrm{w}}^{2}}\ge\left(\left(\frac{\sqrt{2}-1}{\sqrt{\left|V\right|}-1}\right)\left|E\right|\right)^{2}\ge\left(\sqrt{2}-1\right)^{2}\frac{\left|E\right|^{2}}{\left|V\right|}.
\]

\end{IEEEproof}
\bibliographystyle{IEEEtran}
\bibliography{bibtextLDPCScaling}

\begin{thebibliography}{10}
\providecommand{\url}[1]{#1}
\csname url@samestyle\endcsname
\providecommand{\newblock}{\relax}
\providecommand{\bibinfo}[2]{#2}
\providecommand{\BIBentrySTDinterwordspacing}{\spaceskip=0pt\relax}
\providecommand{\BIBentryALTinterwordstretchfactor}{4}
\providecommand{\BIBentryALTinterwordspacing}{\spaceskip=\fontdimen2\font plus
\BIBentryALTinterwordstretchfactor\fontdimen3\font minus
  \fontdimen4\font\relax}
\providecommand{\BIBforeignlanguage}[2]{{%
\expandafter\ifx\csname l@#1\endcsname\relax
\typeout{** WARNING: IEEEtran.bst: No hyphenation pattern has been}%
\typeout{** loaded for the language `#1'. Using the pattern for}%
\typeout{** the default language instead.}%
\else
\language=\csname l@#1\endcsname
\fi
#2}}
\providecommand{\BIBdecl}{\relax}
\BIBdecl

\bibitem{GallagerLDPC}
R.~Gallager, ``Low-density parity-check codes,'' \emph{{IRE} Trans.\ Info.\
  Theory}, vol.~8, no.~1, pp. 21--28, 1962.

\bibitem{OswaldCapacityApproachingLDPCErasureChannel}
P.~Oswald and A.~Shokrollahi, ``Capacity-achieving sequences for the erasure
  channel,'' \emph{IEEE Trans.\ Info.\ Theory}, vol.~48, no.~12, pp.
  3017--3028, Dec. 2002.

\bibitem{Thompson}
\BIBentryALTinterwordspacing
C.~D. Thompson, ``Area-time complexity for {VLSI},'' in \emph{Proceedings of
  the Eleventh Annual ACM Symposium on Theory of Computing}, ser. STOC
  '79.\hskip 1em plus 0.5em minus 0.4em\relax New York, NY, USA: ACM, 1979, pp.
  81--88. [Online]. Available: \url{http://doi.acm.org/10.1145/800135.804401}
\BIBentrySTDinterwordspacing

\bibitem{groverFundamental}
P.~Grover, A.~Goldsmith, and A.~Sahai, ``Fundamental limits on the power
  consumption of encoding and decoding,'' in \emph{Proc. 2012 IEEE Int. Symp.
  Info. Theory}, 2012, pp. 2716--2720.

\bibitem{BlakeKschischangFundamentalLowerBoundArxiv}
\BIBentryALTinterwordspacing
C.~G. Blake and F.~R. Kschischang, ``Energy consumption of {VLSI} decoders,''
  \emph{CoRR}, vol. abs/1412.4130, 2014. [Online]. Available:
  \url{http://arxiv.org/abs/1412.4130}
\BIBentrySTDinterwordspacing

\bibitem{GanesonGroverLDPCLowerBound}
K.~Ganesan, P.~Grover, and A.~Goldsmith, ``How far are {LDPC} codes from
  fundamental limits on total power consumption?'' in \emph{50th Ann.\ Allerton
  Conf.\ Commun., Control, and Comput.}, Monticello, IL, 2012, pp. 671--678.

\bibitem{FiedlerBisectionWidthLowerBound}
M.~Fiedler, ``A property of the eigenvectors of non-negative symmetric matrices
  and its application to graph theory,'' \emph{Czechoslovak Mathematical},
  vol.~25, pp. 619--633, 1975.

\bibitem{BezrukovSpectralLowerBounds}
S.~Bezrukov, R.~Elsasser, B.~Monien, R.~Preis, and J.-P. Tillich, ``New
  spectral lower bounds on the bisection width of graphs,'' \emph{Theoretical
  Computer Science}, vol. 320, pp. 155--174, Mar. 2004.

\bibitem{DiazBisectionWidthBoundsRegularGraphs}
J.~Diaz, M.~J. Serna, and N.~C. Wormald, ``Bounds on the bisection width for
  random d-regular graphs,'' \emph{Theoretical Computer Science}, vol. 382, pp.
  120--130, 2007.

\bibitem{Garey1976237}
\BIBentryALTinterwordspacing
M.~Garey, D.~Johnson, and L.~Stockmeyer, ``Some simplified {NP}-complete graph
  problems,'' \emph{Theoretical Computer Science}, vol.~1, no.~3, pp. 237 --
  267, 1976. [Online]. Available:
  \url{http://www.sciencedirect.com/science/article/pii/0304397576900591}
\BIBentrySTDinterwordspacing

\bibitem{1023755}
J.~Thorpe, ``Design of {LDPC} graphs for hardware implementation,'' in
  \emph{Proceedings of 2002 IEEE International Symposium on Information
  Theory}, 2002, p. 483.

\bibitem{groverOverDesigning}
K.~Ganesan, P.~Grover, and J.~Rabaey, ``The power cost of over-designing
  codes,'' in \emph{Proc.\ 2011 IEEE Workshop Signal Proc.\ Sys.}, 2011, pp.
  128--133.

\bibitem{douglasb.west2001}
D.~B. West, \emph{Introduction to Graph Theory}, 2nd~ed.\hskip 1em plus 0.5em
  minus 0.4em\relax Prentice Hall, 2001.

\bibitem{Richardson:2008:MCT:1795974}
T.~Richardson and R.~Urbanke, \emph{Modern Coding Theory}.\hskip 1em plus 0.5em
  minus 0.4em\relax New York, NY, USA: Cambridge University Press, 2008.

\bibitem{Sason}
I.~Sason, ``On universal properties of capacity-approaching {LDPC} code
  ensembles,'' \emph{IEEE Trans.\ Info.\ Theory}, vol.~55, no.~7, pp. 1--2,
  Jul. 2009.

\bibitem{GallagerBook}
R.~G. Gallager, \emph{Information Theory and Reliable Communication}.\hskip 1em
  plus 0.5em minus 0.4em\relax New York, NY, USA: John Wiley \& Sons, Inc.,
  1968.

\bibitem{Strassen}
V.~Strassen, ``Asymptotische abschatzungen in {S}hannons
  {I}nformationstheorie,'' in \emph{Trans.\ 3rd Prague Conf.\ Info.\ Theory,
  Statist.\ Decision Functions, Random Proc.}\hskip 1em plus 0.5em minus
  0.4em\relax Prague: Pub. House Czechoslovak Acad.\ Sciences, 1962, pp.
  689--723.

\bibitem{RothEtAl}
C.~Roth, A.~Cevrero, C.~Studer, Y.~Leblebici, and A.~Burg, ``Area, throughput,
  and energy-efficiency trade-offs in the {VLSI} implementation of {LDPC}
  decoders,'' in \emph{2011 IEEE International Symposium on Circuits and
  Systems (ISCAS)}, May 2011, pp. 1772--1775.

\end{thebibliography}

\end{document}